\documentclass[preprint]{aastex63}
\usepackage{xspace} 
\usepackage{graphicx} 
\usepackage{natbib}
\usepackage{amsmath}
\usepackage{afterpage}
\newcommand{\y}[1]{\ensuremath{y_{#1}}\xspace}

\newcommand{\gsim}{\ensuremath{\,\gtrsim\,}\xspace}

\newcommand{\gl}{\ensuremath{\ell}\xspace}
\newcommand{\gb}{\ensuremath{{\it b}}\xspace}

\newcommand{\kms}{\ensuremath{\,{\rm km\,s^{-1}}}\xspace}

\newcommand{\m}{\ensuremath{\,{\rm m}}\xspace}

\newcommand{\K}{\ensuremath{\,{\rm K}}\xspace}

\newcommand{\mhz}{\ensuremath{\,{\rm MHz}}\xspace}
\newcommand{\ghz}{\ensuremath{\,{\rm GHz}}\xspace}

\newcommand{\arcmper}{\ensuremath{{^{\prime}}}}
\newcommand{\arcsper}{\ensuremath{{{^{\prime\prime}}}}}

\newcommand{\h}[1]{\ensuremath{^{#1}{\rm H}}\xspace}
\newcommand{\hi}{H\,{\sc i}}
\newcommand{\hii}{H\,{\sc ii}}

\defcitealias{wenger2019a}{Paper~I}
\newcommand{\vect}[1]{\boldsymbol{#1}}

\submitjournal{The Astrophysical Journal Supplement Series}
\received{January 4, 2021}
\revised{February 24, 2021}
\accepted{March 15, 2021}

\shorttitle{SHRDS II: Full Catalog}
\shortauthors{Wenger et al.}

\begin{document}

\title{The Southern \hii\ Region Discovery Survey II: The Full Catalog}

\author[0000-0003-0640-7787]{Trey V. Wenger} 
\affiliation{Dominion Radio Astrophysical Observatory, Herzberg Astronomy and
  Astrophysics Research Centre, National Research Council, P.O. Box 248,
  Penticton, BC V2A 6J9, Canada.}
\email{Trey.Wenger@nrc-cnrc.gc.ca}

\author[0000-0003-0235-3347]{J. R. Dawson}
\affiliation{Department of Physics and Astronomy and MQ Research
  Centre in Astronomy, Astrophysics, and Astrophotonics,
  Macquarie University, NSW 2109, Australia.}
\affiliation{Australia Telescope National Facility, CSIRO Astronomy
  and Space Science, P.O. Box 76, Epping, NSW 1710, Australia.}

\author[0000-0002-6300-7459]{John M. Dickey}
\affiliation{School of Natural Sciences, University of Tasmania,
  Hobart, TAS 7001, Australia.}

\author[0000-0002-1220-2940]{C. H. Jordan}
\affiliation{International Centre for Radio Astronomy Research,
  Curtin University, Bentley, WA 6102, Australia.}
\affiliation{ARC Centre of Excellence for All Sky Astrophysics in 3
  Dimensions (ASTRO 3D), Curtin University, Bentley 6845, Australia}

\author[0000-0003-2730-957X]{N. M. McClure--Griffiths}
\affiliation{Research School of Astronomy and Astrophysics,
  Australian National University, Canberra, ACT 2611, Australia.}

\author[0000-0001-8800-1793]{L. D. Anderson}
\affiliation{Department of Physics and Astronomy, West Virginia
  University, Morgantown, WV 26505, USA.}
\affiliation{Center for Gravitational Waves and Cosmology, West
Virginia University, Morgantown, Chestnut Ridge Research Building, 
Morgantown, WV 26505, USA.}
\affiliation{Adjunct Astronomer at the Green Bank Observatory, 
  P.O. Box 2, Green Bank, WV 24944, USA.}

\author[0000-0002-7045-9277]{W. P. Armentrout}
\affiliation{Green Bank Observatory, P.O. Box 2, Green Bank, WV 24944, USA.}

\author[0000-0002-2465-7803]{Dana S. Balser}
\affiliation{National Radio Astronomy Observatory, 520 Edgemont Road,
  Charlottesville, VA 22903, USA.}

\author[0000-0003-4866-460X]{T. M. Bania}
\affiliation{Institute for Astrophysical Research, Astronomy 
  Department, Boston University, 725 Commonwealth Ave., Boston, MA 
  02215, USA.}

\begin{abstract}
  The Southern \hii\ Region Discovery Survey (SHRDS) is a 900 hour
  Australia Telescope Compact Array 4--10\ghz\ radio continuum and
  radio recombination line (RRL) survey of Galactic \hii\ regions and
  infrared-identified \hii\ region candidates in the southern sky. For
  this data release, we reprocess all previously published SHRDS data
  and include an additional \({\sim}450\) hours of observations. The
  search for new \hii\ regions is now complete over the range
  \(259^\circ < \gl < 346^\circ, |\gb| < 4^\circ\) for \hii\ region
  candidates with predicted \(6\ghz\) continuum peak brightnesses
  \(\gsim30\,\text{mJy beam}^{-1}\). We detect radio continuum
  emission toward 730 targets altogether including previously known
  nebulae and \hii\ region candidates. By averaging \({\sim}18\) RRL
  transitions, we detect RRL emission toward 206 previously known
  \hii\ regions and 436 \hii\ region candidates. Including the
  northern sky surveys, over the last decade the \hii\ Region
  Discovery Surveys have more than doubled the number of known
  Galactic \hii\ regions. The census of \hii\ regions in the
  \textit{WISE} Catalog of Galactic \hii\ Regions is now complete for
  nebulae with 9\ghz\ continuum flux densities
  \(\gsim250\,\text{mJy}\). We compare the RRL properties of the newly
  discovered SHRDS nebulae with those of all previously known
  \hii\ regions. The median RRL full-width at half-maximum line width
  of the entire \textit{WISE} Catalog \hii\ region population
  is \(23.9\kms\) and is consistent between Galactic quadrants. The
  observed Galactic longitude-velocity asymmetry in the population of
  \hii\ regions probably reflects underlying spiral structure in the
  Milky Way.
\end{abstract}


\section{Introduction}

\hii\ regions are the zones of ionized gas surrounding young, massive
stars. These nebulae are the archetypical tracer of spiral arms and
gas-phase metallicity structure in galaxies. Their physical properties
(e.g., electron density, electron temperature, metallicity) inform our
understanding of high-mass star formation
\citep[e.g.,][]{churchwell2002}, the interstellar medium \citep[ISM;
  e.g.,][]{luisi2016}, and Galactic chemical evolution
\citep[e.g.,][]{wenger2019b}. A complete census of Galactic
\hii\ regions will place constraints on models of Milky Way formation
and evolution, but a lack of sensitive surveys in the southern sky has
left this census incomplete.

Dedicated searches for Galactic \hii\ regions began nearly seven
decades ago with the photographic plate surveys of
\citet{sharpless1953} and \citet{sharpless1959} in the northern
hemisphere and \citet{gum1955} and \citet{rodgers1960} in the southern
hemisphere. The prediction \citep{kardashev1959} and discovery
\citep{hoglund1965a,hoglund1965b} of radio recombination lines (RRLs)
provided a new, extinction-free spectroscopic tracer of Galactic
\hii\ regions. Over the next several decades, RRL surveys discovered
hundreds of new nebulae, the majority of which are optically obscured
\citep[e.g.,][]{reifenstein1970,wilson1970,downes1980,caswell1987,lockman1989,lockman1996}.
See \citet[][the Bright Catalog; hereafter, Paper I]{wenger2019a} for
a brief review of the history of \hii\ region RRL surveys.

The \textit{Wide-field Infrared Survey Explorer (WISE)} Catalog of
Galactic \hii\ Regions \citep[][hereafter, the \textit{WISE}
  Catalog]{anderson2014} requires the detection of recombination line
emission, such as H\(\alpha\) or RRL emission, to classify a nebula as
a known \hii\ region. This is a conservative definition of an
\hii\ region. Other studies have less restrictive criteria for
\hii\ region classification. For example, the presence of ionized gas
in nebulae has been inferred from radio continuum spectral energy
distributions \citep[e.g.,][]{becker1994} and/or infrared colors
\citep[e.g.,][]{wood1989,white1991}, although these techniques
sometimes suffer from contamination by other Galactic objects such as
planetary nebulae or high-mass evolved stars
\citep[e.g.,][]{leto2009}. Recombination line emission is unambiguous
evidence for the presence of thermally-emitting plasma.

The Southern \hii\ Region Discovery Survey (SHRDS) is the final
component of the \hii\ Region Discovery Survey
\citep[HRDS;][]{bania2010}. The original Green Bank Telescope (GBT)
HRDS discovered 602 hydrogen RRLs toward 448 \hii\ region candidates
in the zone \(67^\circ \leq \gl \leq 343^\circ; |\gb| \leq
1^\circ\). These candidates were selected based on their spatially
coincident radio continuum and 24 \(\mu\)m emission
\citep{anderson2011}. Subsequent HRDS papers derived kinematic
distances \citep{anderson2012} and characterized the helium and carbon
RRLs \citep{wenger2013} for a subset of these nebulae.  Additional
surveys made with the Arecibo Telescope \citep{bania2012} and the GBT
\citep{anderson2015b,anderson2018} extended the HRDS. Altogether, the
HRDS discovered 887 new Galactic \hii\ regions in the northern sky.

The SHRDS extends the HRDS to the southern sky. We use the Australia
Telescope Compact Array (ATCA) to search for 4--10\ghz\ radio
continuum and RRL emission toward infrared-identified \hii\ region
candidates in the 3rd and 4th Galactic quadrants. In the pilot survey
\citep{brown2017} and first data release \citepalias{wenger2019a}, we
found RRL emission toward 295 new Galactic \hii\ regions, nearly
doubling the number of known nebulae in the survey zone.

The Full Catalog presented in this paper is the final SHRDS data
release. We add an additional \({\sim}450\) hours of observations not
included in \citetalias{wenger2019a} and also reanalyze all the
previous SHRDS data. By reprocessing all of the data in a uniform and
consistent manner, we minimize any systematic discrepancies between
data releases. This data release supersedes the Pilot survey
\citep{brown2017} and \citetalias{wenger2019a}. Although the data
reduction and analysis steps are similar, we make several changes to
improve the data quality and maximize our detection rate. Furthermore,
we include here intermediate data products, such as the properties of
individual RRL transitions, which can be used to study specific
nebulae in detail.

\section{Target Sample}

We select the SHRDS \hii\ region candidate targets from the
\textit{WISE} Catalog. In \citetalias{wenger2019a}, we targeted
nebulae with predicted 6\ghz\ continuum brightnesses \(\gsim
60\,\text{mJy beam}^{-1}\) based on extrapolated Sydney University
Molonglo Sky Survey (SUMSS) 843\mhz\ flux densities and an assumed
optically thin spectral index of \(\alpha=-0.1\). For the Full
Catalog, we add those \textit{WISE} Catalog objects with predicted
brightnesses between \(30\) and \(60\,\text{mJy beam}^{-1}\) and
infrared diameters smaller than \(5\arcmper\). The optically thin
assumption is likely invalid at frequencies \(\lesssim 1\ghz\), and
indeed we find a scatter of \({\sim}100\%\) between the extrapolated
and measured 6\ghz flux densities of nebulae in the Bright Catalog
\citepalias[see][]{wenger2019a}. Objects in the \textit{WISE} Catalog
that do not meet our brightness criterion may be nebulae ionized by
lower-mass stars (B-stars), very distant, or optically thick.

As both a test of our experiment and to improve the accuracy and
reliability of the previous single dish RRL detections, we also
observe 175 previously known \hii\ regions. These nebulae have
previous RRL detections made with the Parkes telescope
\citep{caswell1987}, the GBT as part of the HRDS, or the Jansky Very
Large Array \citep[VLA;][]{wenger2019b}. The \citet{caswell1987} RRL
survey has an angular resolution of \({\sim}4.4\arcmper\), which is
insufficient to disentangle the RRL emission from multiple
\hii\ regions in confusing fields. The GBT HRDS and VLA data overlap
with the higher frequency end of the SHRDS (\({\sim}\)8--10\ghz) and
have comparable (\({\sim}90\arcsper\) for GBT HRDS) or finer
(\({\sim}10\arcsper\) for VLA) angular resolution.  We test our
experiment by comparing the RRL properties (e.g., line width) measured
by the SHRDS with those previous detections.

The target list for the full SHRDS includes 435 \hii\ region
candidates and 175 previously known \hii\ regions. The ATCA has a
large field of view (primary beam half-power beam width, HPBW,
\({\sim}4\arcmper\) at \(8\ghz\)) so multiple \textit{WISE} catalog sources
can appear within a single pointing. Table~\ref{tab:fields} gives
information about the \hii\ regions and \hii\ region candidates in
each SHRDS field: the field name, the field center position, and every
\textit{WISE} catalog source within that field. For each source, we
list: the \textit{WISE} catalog source name, the source type (``K''
for previously known \hii\ region; ``C'' for \hii\ region candidate;
``Q'' for radio-quiet \hii\ region candidate, which lacks detected
radio continuum emission in extant surveys; and ``G'' for a candidate
associated with a known group of \hii\ regions), the \textit{WISE}
infrared position, the \textit{WISE} infrared radius, and the
reference to the previous non-SHRDS RRL detection, if any.

\section{Observations, Data Reduction \& Analysis}

The observing strategy and data processing procedure for the Full
Catalog are similar to that of \citetalias{wenger2019a}. We used the
ATCA C/X-band receiver and Compact Array Broadband Backend (CABB) to
simultaneously observe 4--8\ghz\ radio continuum emission and 20
hydrogen RRL transitions in two orthogonal linear polarizations (note
that \citetalias{wenger2019a} incorrectly states that circular
polarizations were observed). Our observations took place between June
2015 and January 2019 with a total of 900 hours of telescope time
split equally between the H75 and H168 antenna configurations. We
observed each field for a total of \({\sim}30\) to \({\sim}90\)
minutes in \({\sim}5\) to \({\sim}10\) minute snapshots spread over
\({\sim}9\) hours in hour angle. See Tables~\ref{tab:observations} and
\ref{tab:backend} for a summary of the observations and spectral
window configuration, respectively.

The ATCA with CABB is an excellent tool for discovering RRLs toward
Galactic \hii\ regions. In our hybrid H75/H168 data, the synthesized
HPBW (\({\sim}90\arcsper\) at \({\sim}8\ghz\)) is well-matched to the
typical classical \hii\ region diameter (90\arcsper is equivalent to a
10 pc diameter at 20 kpc distance). The compact antenna configurations
yield a good surface brightness sensitivity for resolved nebulae, and
the large collecting area yields a good point-source sensitivity for
unresolved nebulae. The large bandwidth and flexible backend allow for
the simultaneous observation of 20 RRL transitions, which we average
to improve the spectral sensitivity. See \citetalias{wenger2019a} for
representative SHRDS images and RRL spectra.

We use the Wenger Interferometry Software Package
\citep[WISP;][]{wisp} to calibrate, reduce, and analyze the SHRDS
data. See \citetalias{wenger2019a} for details regarding the
calibration of the SHRDS data. WISP is a wrapper for the
\textit{Common Astronomy Software Applications} package
\citep[CASA;][]{mcmullin2007}. WISP uses automatic flagging and
\textit{CLEAN} region identification to reduce the burden of
interferometric data processing, which can be extremely time
intensive. WISP is well-tested on both ATCA and VLA radio continuum
and spectral line data \citep{wenger2019b,wenger2019a}.

For each observed field the WISP imaging pipeline produces the
following data products: a 4\ghz\ bandwidth multi-scale,
multi-frequency synthesis (MS-MFS) continuum image, sixteen
256\mhz\ bandwidth MS-MFS continuum images covering the full
4\ghz\ bandwidth, a MS-MFS continuum image of each 64\mhz\ bandwidth
spectral line window, and a multi-scale data cube of each spectral
line window. The first and last of the 256\mhz\ bandwidth continuum
images are compromised by radio-frequency interference (RFI) and
band-edge effects and are typically unusable.

Unlike \citetalias{wenger2019a}, we image each spectral line data cube
at its native spectral resolution. Since the RRLs span almost a factor
of two in frequency, the velocity resolution of our spectral line
windows varies by nearly a factor of two as well. The spectra must be
sampled or re-gridded on a common velocity axis in order to average
the individual RRL spectra and create the
\(\langle\text{Hn}\alpha\rangle\) average spectrum. We improve the
sensitivity of our average spectra by smoothing the higher resolution
spectra to match the lowest resolution observed.
\citetalias{wenger2019a} uses linear interpolation (without smoothing)
to re-grid the spectra. Here we use sinc interpolation to both smooth
and re-grid the RRL spectra to a common velocity axis (see
Appendix~\ref{app:sinc}). This method results in a \({\sim}20\%\)
sensitivity improvement in the \(\langle\text{Hn}\alpha\rangle\)
average spectra compared to the \citetalias{wenger2019a} analysis.

\clearpage
\begin{longrotatetable}
\begin{deluxetable*}{lccccccrl}
\centering
\tablewidth{0pt}
\tabletypesize{\scriptsize}
\tablecaption{Full Catalog Fields and Targets\label{tab:fields}}
\tablehead{
\colhead{Field} & \colhead{RA}         & \colhead{Dec.}       & \colhead{Target} & \colhead{Catalog\tablenotemark{a}} & \colhead{RA}         & \colhead{Dec.}       & \colhead{\(R_{\rm IR}\)}   & \colhead{Author} \\
\colhead{}      & \colhead{J2000}      & \colhead{J2000}      & \colhead{}       & \colhead{}                         & \colhead{J2000}      & \colhead{J2000}      & \colhead{(arcsec)}         & \colhead{}      \\
\colhead{}      & \colhead{(hh:mm:ss)} & \colhead{(dd:mm:ss)} & \colhead{}       & \colhead{}                         & \colhead{(hh:mm:ss)} & \colhead{(dd:mm:ss)} & \colhead{}                 & \colhead{}      
}
\startdata
ch1 & 07:30:05.8 & $-18$:32:27.2 & G233.676$-$00.186 & Q & 07:29:56.6 & $-18$:27:50.1 & $59.11$ &  \\
 & & & G233.753$-$00.193 & K & 07:30:04.6 & $-18$:32:03.9 & $311.06$ & B11  \\
 & & & G233.830$-$00.180 & G & 07:30:16.9 & $-18$:35:44.5 & $114.85$ &  \\
ch4 & 08:20:56.3 & $-36$:12:32.0 & G254.682+00.220 & K & 08:20:55.5 & $-36$:13:09.9 & $385.32$ & CH87  \\
shrds027 & 08:23:21.5 & $-40$:39:59.7 & G258.608$-$01.925 & Q & 08:23:21.7 & $-40$:39:57.7 & $136.61$ &  \\
shrds029 & 08:26:13.4 & $-40$:46:45.4 & G259.013$-$01.546 & Q & 08:26:13.5 & $-40$:46:45.0 & $68.08$ &  \\
 & & & G259.057$-$01.544 & C & 08:26:22.1 & $-40$:48:49.5 & $103.98$ &  \\
 & & & G259.086$-$01.612 & C & 08:26:09.9 & $-40$:52:36.2 & $103.98$ &  \\
shrds030 & 08:26:18.8 & $-40$:48:36.5 & G259.013$-$01.546 & Q & 08:26:13.5 & $-40$:46:45.0 & $68.08$ &  \\
 & & & G259.057$-$01.544 & C & 08:26:22.1 & $-40$:48:49.5 & $103.98$ &  \\
 & & & G259.086$-$01.612 & C & 08:26:09.9 & $-40$:52:36.2 & $103.98$ &  \\
\enddata
\tablecomments{This table is available in its entirety in a machine-readable form in the online journal. A portion is shown here for guidance regarding its form and content.}
\tablenotetext{a}{The \textit{WISE} Catalog designation: ``K'' is known \hii\ region, ``C'' is \hii\ region candidate, ``Q'' is radio-quiet \hii\ region candidate, and ``G'' is \hii\ candidate associated with an \hii\ region group.}
\tablerefs{(GBT HRDS) \citet{anderson2011,anderson2015b,anderson2018}; (B11) \citet{balser2011}; (CH87) \citet{caswell1987}; (D80) \citet{downes1980}; (L89) \citet{lockman1989}; (Q06) \citet{quireza2006a}; (S04) \citet{sewilo2004}; (W70) \citet{wilson1970}}
\end{deluxetable*}
\end{longrotatetable}

\clearpage

\begin{table}
  \centering
  \caption{Observation Summary\label{tab:observations}}
  \begin{tabular}{ll}
    \hline
    \hline
    Observing Dates & 2015$-$07$-$24 to 2019$-$01$-$25 \\
    Observing Time & 900 hr \\
    Primary Calibrators & 0823$-$500, 1934$-$638 \\
    Secondary Calibrators & 0906$-$47, 1036$-$52, j1322$-$6532, 1613$-$586 \\
    & 1714$-$397, 1714$-$336, 1829-207 \\
    \hline
  \end{tabular}
\end{table}

In the Full Catalog, we use the polarization data to generate
full-Stokes (IQUV) MS-MFS images of the 4\ghz\ bandwidth continuum
image and each 256\mhz\ bandwidth continuum image. To calibrate the
polarization data, we assume that the primary calibrator 1934$-$638 is
unpolarized and any observed polarized emission is due to instrumental
leakage. We use the 1934$-$638 data to derive the instrumental
polarization leakage corrections. After applying these leakage
corrections to the secondary calibrators, we derive the Stokes
polarization fractions (\(Q/I\), \(U/I\), and \(V/I\)) for each
secondary calibrator, update the secondary calibrator flux model with
these polarization fractions, and recompute the complex gain
calibration tables.  Finally, we apply all of the calibration tables,
including the polarization leakage calibration tables, to the science
targets. This polarization calibration prescription is included in the
latest WISP release.

\begin{deluxetable*}{lcccccc}
\centering
\tablewidth{0pt}
\tabletypesize{\scriptsize}
\tablecaption{Spectral Window Configuration\label{tab:backend}}
\tablehead{
  \\
  \colhead{Window} & \colhead{Center Freq.} & \colhead{Bandwidth} & \colhead{Channels} & \colhead{Channel Width} & \colhead{RRL} & \colhead{Rest Freq.} \\
  \colhead{}       & \colhead{(MHz)}  & \colhead{(MHz)}     & \colhead{}         & \colhead{(kHz)}   & \colhead{}    & \colhead{(MHz)}
}
\startdata
0      & 4545         & 256       & 4        & 64000         & \dots & \dots \\
1      & 4801         & 256       & 4        & 64000         & \dots & \dots \\
2      & 5057         & 256       & 4        & 64000         & \dots & \dots \\
3      & 5313         & 256       & 4        & 64000         & \dots & \dots \\
4      & 5569         & 256       & 4        & 64000         & \dots & \dots \\
5      & 5825         & 256       & 4        & 64000         & \dots & \dots \\
6      & 6081         & 256       & 4        & 64000         & \dots & \dots \\
7      & 6337         & 256       & 4        & 64000         & \dots & \dots \\
8      & 7580         & 256       & 4        & 64000         & \dots & \dots \\
9      & 7836         & 256       & 4        & 64000         & \dots & \dots \\
10     & 8092         & 256       & 4        & 64000         & \dots & \dots \\
11     & 8348         & 256       & 4        & 64000         & \dots & \dots \\
12     & 8604         & 256       & 4        & 64000         & \dots & \dots \\
13     & 8860         & 256       & 4        & 64000         & \dots & \dots \\
14     & 9116         & 256       & 4        & 64000         & \dots & \dots \\
15     & 9372         & 256       & 4        & 64000         & \dots & \dots \\
16 & 4609 & 64 & 2049 & 31.25 & H112\(\alpha\) & 4618.790 \\
17 & 4737 & 64 & 2049 & 32.25 & H111\(\alpha\) & 4744.184 \\
18 & 4865 & 64 & 2049 & 31.25 & H110\(\alpha\) & 4874.158 \\
19 & 4993 & 64 & 2049 & 31.25 & H109\(\alpha\) & 5008.924 \\
20 & 5153 & 64 & 2049 & 31.25 & H108\(\alpha\) & 5148.704 \\
21 & 5281 & 64 & 2049 & 31.25 & H107\(\alpha\) & 5293.733 \\
22 & 5441 & 64 & 2049 & 31.25 & H106\(\alpha\) & 5444.262 \\
23 & 5601 & 64 & 2049 & 31.25 & H105\(\alpha\) & 5600.551 \\
24 & 5761 & 64 & 2049 & 31.25 & H104\(\alpha\) & 5762.881 \\
25 & 5921 & 64 & 2049 & 31.25 & H103\(\alpha\) & 5931.546 \\
26 & 6113 & 64 & 2049 & 31.25 & H102\(\alpha\) & 6106.857 \\
27 & 6305 & 64 & 2049 & 31.25 & H101\(\alpha\) & 6289.145 \\
28 & 6465 & 64 & 2049 & 31.25 & H100\(\alpha\) & 6478.761 \\
29 & 7548 & 64 & 2049 & 31.25 & H95\(\alpha\) & 7550.616 \\
30 & 7804 & 64 & 2049 & 31.25 & H94\(\alpha\) & 7792.872 \\
31 & 8060 & 64 & 2049 & 31.25 & H93\(\alpha\) & 8045.604 \\
32 & 8316 & 64 & 2049 & 31.25 & H92\(\alpha\) & 8309.384 \\
33 & 8572 & 64 & 2049 & 31.25 & H91\(\alpha\) & 8584.823 \\
34 & 9180 & 64 & 2049 & 31.25 & H89\(\alpha\) & 9173.323 \\
35 & 8500 & 64 & 2049 & 31.25 & H88\(\alpha\) & 9487.823 \\
\enddata
\end{deluxetable*}

We do not perform any \textit{uv}-tapering when imaging SHRDS Full
Catalog fields. Tapering can improve surface brightness sensitivity at
the expense of angular resolution. In \citetalias{wenger2019a}, we
\textit{uv}-tapered all data to a \({\sim}100\arcsper\) angular
resolution and then smoothed these images to a common angular
resolution. By the convolution theorem, smoothing by convolving the
images with a Gaussian kernel is equivalent to \textit{uv}-tapering by
weighting the \textit{uv}-data with a Gaussian function
\citep{oppenheim1975}. Therefore, we save processing time and disk
space by skipping the \textit{uv}-tapering step and instead smooth the
non-tapered data to a common angular resolution. In principle, we
might lose sensitivity to diffuse emission that is not bright enough
to be \textit{CLEAN}ed in the non-tapered images. Our surface
brightness sensitivity, however, is excellent due to the compact
antenna configuration and comprehensive \textit{uv}-coverage, and
tests with \textit{uv}-tapering reveal no noticeable improvement in
sensitivity compared to the smoothed images.

\begin{figure*}
  \centering
  \includegraphics[width=\textwidth]{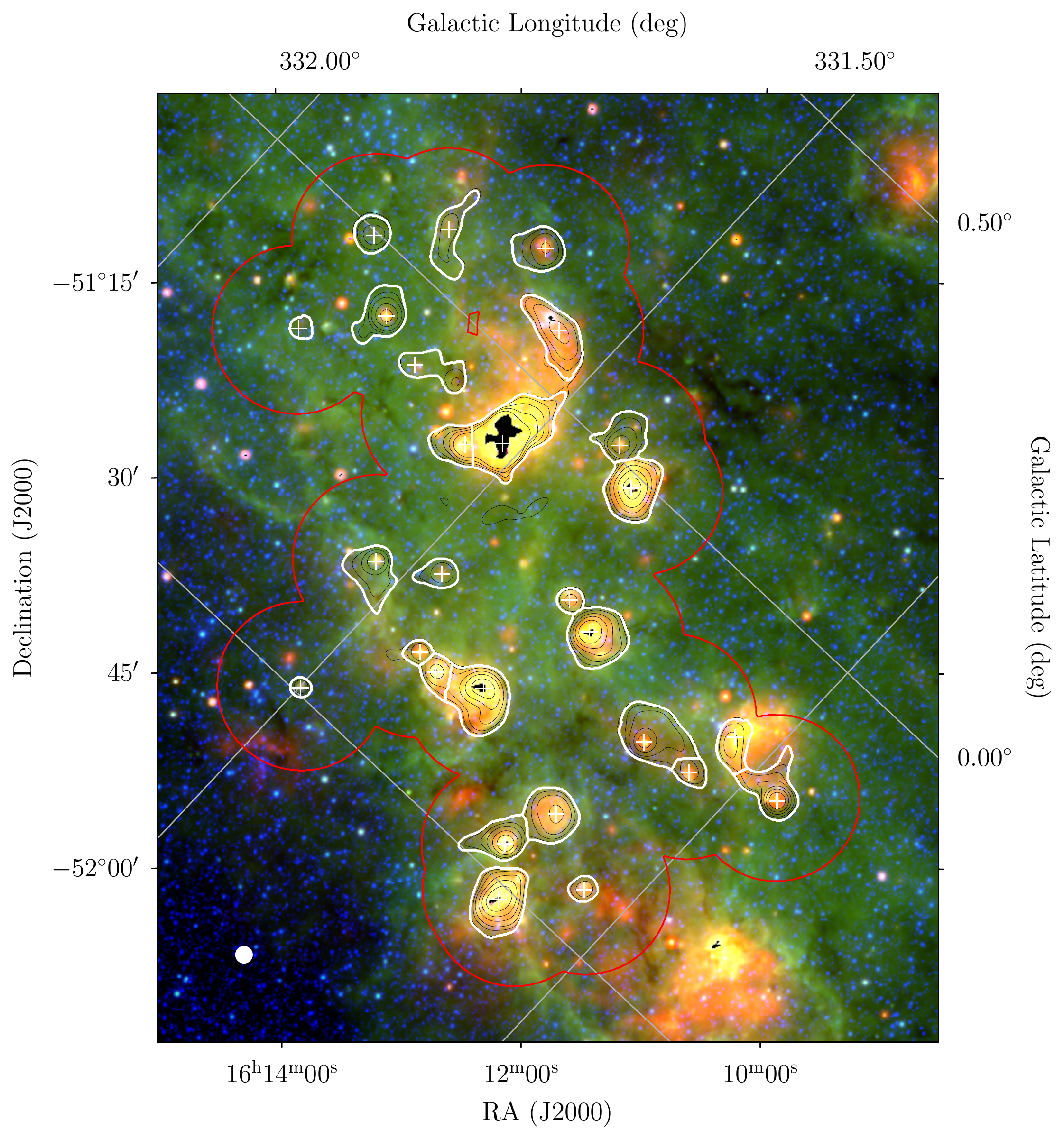}
  \caption{The largest SHRDS radio continuum mosaic, mos040, overlaid
    on 3.4\(\mu\)m (blue), 12\(\mu\)m (green), and 22\(\mu\)m (red)
    data from the \textit{WISE} All-Sky Survey
    \citep{wright2010}. Black regions in the image are saturated in at
    least one \textit{WISE} band. The black contours are the mosaicked
    4\ghz\ bandwidth radio continuum data at 5, 10, 20, 50, 100, and
    200 times the image rms. The red contour represents the 10\%
    primary beam response and the extent of the SHRDS data. Each white
    cross is the location of the brightest pixel associated with a
    \textit{WISE} Catalog source and each white contour is the
    watershed region for that source. The filled white circle in the
    lower left is the smoothed resolution of the radio continuum
    data.}
  \label{fig:mosaic}
\end{figure*}

We extract spectra and measure the total flux densities of the SHRDS
nebulae using the watershed segmentation technique of
\citet{wenger2019b}. Given an image, the watershed segmentation
algorithm identifies all pixels with emission associated with an
emission peak \citep[the ``watershed region;''
  see][]{bertrand2005}. If two or more peaks are close together, then
the algorithm will split the emission into that many non-overlapping
watershed regions. As in \citetalias{wenger2019a}, we first identify
continuum detections as any emission peaks that are (1) brighter than
5 times the rms noise in the 4\ghz\ bandwidth MS-MFS image and (2)
within the circle defined by the \textit{WISE} Catalog infrared
position and radius.  To estimate the rms noise across the image, we
divide the standard deviation of the residual image by the primary
beam response. Hereafter, the location of the emission peak in the
4\ghz\ bandwidth MS-MFS image is called the source position. For each
continuum detection we measure the peak continuum brightness at the
source position in every MS-MFS image. We extract a spectrum from each
data cube at this position as well. The source positions are the
starting locations for the watershed segmentation algorithm. We clip
each MS-MFS image by masking all pixels fainter than 5 times the rms
noise and then apply the watershed algorithm to these clipped
images. Following the procedure in \citet{wenger2019b}, we measure
total continuum flux densities in each MS-MFS image and extract
spectra from every data cube. The total flux density is the sum of the
brightness in each pixel within the watershed region. The watershed
region spectrum is a weighted average of the spectra extracted from
each pixel within the region \citep[see][]{wenger2019b}.

The SHRDS continuum and RRL flux densities are systematically
underestimated. By the nature of an interferometer, we miss flux due
to the central ``hole'' in our uv-coverage. The importance of this
effect depends on the uv-coverage of the data and the source
brightness distribution. For a Gaussian source brightness
distribution, we use equation A6 from \citet{wilner1994} to estimate
that the SHRDS flux densities are reduced by
\({\sim}5\%\). Furthermore, because we use the clipped MS-MFS images
to generate the watershed regions, the measured total flux densities
are further underestimated. For a Gaussian source brightness
distribution, the total flux density is reduced by \({\sim}20\%\) if
the peak brightness is \({\sim}10\) times the rms noise and by
\({\sim}10\%\) if the peak brightness is \({\sim}20\) times the rms
noise. The SHRDS total flux densities may be unreliable if the source
has a complicated emission structure. These effects are eliminated in
the RRL-to-continuum brightness ratio, however, if both the RRL and
continuum flux densities are measured using the same watershed region
in the same data cube.

Many individual SHRDS fields overlap due to the large field of view of
the ATCA. We improve the sensitivity in the overlapped regions by
creating linear mosaics following \citet{cornwell1993}:
\begin{equation}
  I^m(\theta) = \frac{\sum_p A^2_p(\theta)I_p(\theta)}{\sum_p A^2_p(\theta)}, \label{eq:mosaic}
\end{equation}
where \(I^m(\theta)\) is the mosaic datum at sky position \(\theta\),
\(I_p(\theta)\) is the non-primary beam corrected datum at \(\theta\)
in pointing \(p\), \(A_p(\theta)\) is the primary beam weight at
\(\theta\) in \(p\), and the sum is over all overlapping
pointings. Thus, each position of the mosaic is the average of each
individual image weighted by the square of the primary beam.

We create 99 mosaics by combining the images of 329 individual
fields. Table~\ref{tab:mosaics} lists the fields that are used to
create each mosaic. Using equation~\ref{eq:mosaic}, we smooth each
continuum image and RRL data cube to the worst resolution of the
ensemble. We process and analyze each mosaic in the same manner as an
individual field. Our largest mosaic, mos040, is the combination of 30
fields centered near \((\gl, \gb) = (331.5^\circ, -0.15^\circ)\). The
4\ghz\ bandwidth MS-MFS continuum mosaic is shown in
Figure~\ref{fig:mosaic}. The white contours show the watershed regions
of the 27 nebulae identified in this mosaic.

\section{SHRDS: The Full Catalog}

The SHRDS Full Catalog is a compilation of the radio continuum and RRL
properties of every \hii\ region detected in the survey. We observe
609 fields containing 1398 \textit{WISE} catalog sources. The fields
contain 289 previously known \hii\ regions, 554 \hii\ region
candidates, 385 radio-quiet candidates, and 170 candidates that are
associated with an \hii\ region group (see Table~\ref{tab:fields}). We
detect radio continuum emission toward 212 previously known nebulae
and 518 \hii\ region candidates, including 40 radio-quiet and 89 group
candidates. Of those, we detect \(\langle\text{Hn}\alpha\rangle\) RRL
emission toward 204 previously known \hii\ regions and 428
\hii\ region candidates, including 14 radio-quiet and 82 group
candidates. We do not detect some previously known \hii\ regions
because (1) the nebula is larger than the ATCA maximum recoverable
scale (\({\sim}200\arcsper\) at 8\ghz) and/or (2) the nebula is near
the edge of the primary beam and/or (3) the nebula is confused with a
nearby \hii\ region and thus missed by our source identification
method.

\clearpage
\startlongtable
\begin{deluxetable}{ll}
\centering
\tablewidth{0pt}
\tabletypesize{\footnotesize}
\tablecaption{Mosaic Fields\label{tab:mosaics}}
\tablehead{
\colhead{Mosaic} & \colhead{Fields} 
}
\startdata
mos000 & fa489, ch309 \\
mos001 & fa054, gs108, ch307 \\
mos002 & ch282, ch283 \\
mos003 & g351.311+, overlap1 \\
mos004 & ch253, ch254 \\
mos005 & ch246, caswell25 \\
mos006 & ch244, caswell24 \\
mos007 & shrds1240, shrds1241 \\
mos008 & shrds1238, shrds1239 \\
mos009 & ch239, caswell23 \\
mos010 & shrds930, shrds931, shrds932 \\
mos011 & ch234, shrds1237 \\
mos012 & shrds925, g339.717-01.102, shrds923 \\
mos013 & ch230, shrds928, shrds922, shrds920, shrds917 \\
 & shrds918, shrds921 \\
mos014 & shrds913, shrds909, shrds908, shrds1228 \\
mos015 & shrds1223, shrds1225, shrds905, shrds906 \\
mos016 & shrds896, caswell22, ch221, shrds898, shrds899 \\
 & shrds900, shrds901 \\
mos017 & shrds878, shrds875, ch214, shrds884, shrds883 \\
 & shrds880, shrds885, shrds881, shrds886, shrds888 \\
 & shrds1218, shrds890, shrds892, shrds893, shrds1219 \\
 & shrds891, shrds1233 \\
mos018 & shrds869, shrds870 \\
mos019 & shrds866, shrds867, shrds868 \\
mos020 & shrds1205, shrds1209, shrds1212, shrds862, shrds864 \\
mos021 & shrds854, shrds857, shrds1204 \\
mos022 & shrds855, ch209, caswell21 \\
mos023 & shrds835, shrds836 \\
mos024 & shrds833, shrds834 \\
mos025 & caswell20, ch201 \\
mos026 & shrds806, shrds807, shrds810, shrds811 \\
mos028 & shrds800, shrds801, shrds799, ch196, caswell19 \\
 & shrds1200 \\
mos029 & shrds801, shrds800, shrds799, shrds802 \\
mos030 & shrds785, shrds789, shrds793, ch194 \\
mos031 & shrds792, shrds786, shrds788, shrds792, shrds794 \\
mos032 & shrds784, ch191 \\
mos033 & shrds777, ch190 \\
mos034 & shrds773, shrds775 \\
mos035 & shrds771, shrds1181 \\
mos036 & shrds759, shrds764 \\
mos037 & shrds756, shrds763, shrds766 \\
mos038 & shrds753, ch186 \\
mos040 & shrds709, shrds711, shrds714, shrds716, shrds720 \\
 & shrds710, shrds721, shrds717, ch179, ch180 \\
 & shrds725, ch181, shrds731, shrds734, shrds732 \\
 & shrds735, ch182, shrds729, ch185, shrds733 \\
 & shrds736, shrds738, shrds737, shrds736, shrds741 \\
 & shrds743, shrds744, shrds746, shrds743, shrds745 \\
mos041 & shrds731, shrds735 \\
mos042 & shrds723, shrds719 \\
mos043 & shrds1171, shrds1172, ch177 \\
mos044 & shrds1174, shrds1175g \\
mos045 & shrds677, shrds679, shrds686 \\
mos046 & ch171, shrds669 \\
mos047 & shrds651, shrds652, shrds659 \\
mos048 & ch168, shrds656, shrds657 \\
mos049 & shrds634, shrds635, shrds640, shrds639, shrds643 \\
 & shrds648, shrds647, shrds644 \\
mos050 & shrds1167, ch162, caswell18, shrds1168 \\
mos051 & shrds1164, shrds1163, ch156 \\
mos052 & ch154, caswell17 \\
mos053 & caswell16, ch149 \\
mos054 & shrds589, shrds590 \\
mos055 & caswell15, ch145, shrds1151 \\
mos056 & caswell14, ch144 \\
mos057 & shrds1144g, shrds559, shrds1146, shrds1147 \\
mos058 & shrds1133g, shrds1134g, ch138, shrds552, ch139 \\
 & shrds1139, shrds1140 \\
mos059 & ch136, shrds1137 \\
mos060 & shrds535, shrds536, ch131 \\
mos061 & caswell13, ch129 \\
mos062 & shrds501, ch122 \\
mos063 & shrds458, shrds462, shrds463 \\
mos064 & ch114, ch113, shrds440, shrds435, shrds439 \\
 & shrds437 \\
mos065 & shrds1103, shrds432 \\
mos066 & caswell12, ch110, shrds428 \\
mos067 & shrds418, shrds413, shrds1096, shrds1095g, shrds411 \\
 & shrds409 \\
mos068 & shrds386, shrds385, shrds387 \\
mos069 & caswell11, ch96 \\
mos070 & shrds377, g308.033-01 \\
mos071 & caswell10, shrds365, shrds364, ch94 \\
mos072 & shrds308, shrds309 \\
mos073 & shrds296, shrds297 \\
mos074 & shrds294, shrds295 \\
mos075 & ch76, caswell9 \\
mos076 & ch73, caswell7, ch74, caswell8 \\
mos077 & shrds271, shrds273 \\
mos078 & shrds1046, ch71, caswell6 \\
mos079 & shrds253, shrds256, shrds261, shrds258, g298.473+ \\
mos080 & shrds249, ch66, caswell5 \\
mos081 & shrds1041, ch67 \\
mos082 & shrds246, ch65 \\
mos083 & shrds215, g293.936-, shrds219, g293.994- \\
mos084 & shrds207, ch57 \\
mos085 & ch55, caswell4 \\
mos086 & ch52, caswell3, shrds1034 \\
mos087 & shrds1031g, shrds1032 \\
mos088 & shrds1028, shrds191, g290.012- \\
mos089 & ch35, ch36 \\
mos090 & shrds1017, shrds1018g \\
mos091 & ch33, shrds143 \\
mos092 & ch32, caswell2 \\
mos093 & ch19, shrds1007 \\
mos094 & shrds1005, ch17, caswell1 \\
mos095 & shrds088, shrds089, ch13 \\
mos096 & shrds085, ch11, shrds087 \\
mos097 & shrds056, shrds057 \\
mos098 & shrds055, shrds054 \\
mos099 & shrds034, shrds036, shrds037, shrds042 \\
mos100 & shrds029, shrds030 \\
\enddata
\end{deluxetable}

As in \citetalias{wenger2019a}, we attempt to reprocess the SHRDS
pilot survey data \citep{brown2017} with mixed success. Due to the
limited \textit{uv}-coverage of the pilot observations, we are unable
to create images for ten pilot fields. \citet{brown2017} extracted
spectra directly from the visibility data. They detected RRL emission
toward ten sources in these unimaged fields: two previously known
\hii\ regions (G213.833+00.618 and G313.790+00.705) and eight
\hii\ region candidates (G230.354$-$00.597, G290.012$-$00.867,
G290.323$-$02.984, G290.385$-$01.042, G290.674$-$00.133,
G291.596$-$00.239, G295.275$-$00.255, G300.972$+$00.994). The RRL data
for these nebulae are included in \citet{brown2017} and
\citetalias{wenger2019a}. These RRL detections increase the number of
SHRDS RRL detections toward previously known and \hii\ region
candidates to 206 and 436, respectively.

The Full Catalog data products supersede the previous Bright Catalog
in \citetalias{wenger2019a}. Due to the differences in the source
selection and data analysis procedures, there are 29 Bright Catalog
continuum detections and 22 Bright Catalog
\(\langle\text{Hn}\alpha\rangle\) RRL detections that are not in the
Full Catalog. Each of these detections is in confusing fields and our
reprocessing has associated the emission with a different, nearby
\textit{WISE} Catalog object.

\subsection{Continuum Catalog}

The continuum properties of the SHRDS nebulae are summarized in
Table~\ref{tab:unsmoothed_continuum} (non-smoothed) and
Table~\ref{tab:continuum} (smoothed). For each source, we list the
\textit{WISE} Catalog name, position of the continuum emission peak,
field name, and quality factor, QF. Then, for each of the continuum
spectral window MS-MFS images and the combined 4\ghz\ bandwidth MS-MFS
image, we list the image synthesized frequency, \(\nu\), bandwidth,
\(\Delta\nu\), synthesized beam area, source peak continuum
brightness, \(S_{C, P}\), watershed region area, and source total
continuum flux density, \(S_{C, T}\). The total flux density is
measured within the watershed region, and the uncertainty is derived
following Equation~1 in \citet{wenger2019b}.

The QF is a qualitative assessment of the reliability of the measured
source continuum properties. The highest quality detections, QF A, are
unresolved, isolated, and near the center of the field. Intermediate
quality detections, QF B, are (1) resolved yet compact with a single
emission peak, (2) slightly confused, such that its watershed region
borders another source's watershed region, and/or (3) located
off-center in the field. The worst quality detections, QF C, have
unreliable continuum properties because they are (1) resolved and the
emission is not compact with a single, distinct peak, (2) severely
confused with other sources, and/or (3) located near the edge of the
field. We recommend using only the QF A and B data in subsequent
analyses of these data, but we include the QF C data for completeness.

A \textit{WISE} Catalog source can appear in
Tables~\ref{tab:unsmoothed_continuum} and \ref{tab:continuum} multiple
times if it is detected in multiple fields and/or mosaics. Each
detection in a non-mosaic field is an independent measurement of the
source continuum properties. The detection with the best quality
factory should be used for subsequent analyses.

\begin{longrotatetable}
\begin{deluxetable*}{lccccDcDD@{$\,\pm\,$}DDD@{$\,\pm\,$}D}
\centering
\tablewidth{0pt}
\tabletypesize{\scriptsize}
\tablecaption{Un-smoothed Continuum Detections\label{tab:unsmoothed_continuum}}
\tablehead{
\colhead{Source} & \colhead{RA}         & \colhead{Dec.}       & \colhead{Field} & \colhead{QF\tablenotemark{a}} & \multicolumn{2}{c}{$\nu$} & \colhead{$\Delta\nu$} & \multicolumn{2}{c}{Beam}         & \multicolumn{4}{c}{$S_{C,P}$}         & \multicolumn{2}{c}{Region}       & \multicolumn{4}{c}{$S_{C,T}$} \\
\colhead{}       & \colhead{J2000}      & \colhead{J2000}      & \colhead{}      & \colhead{}                    & \multicolumn{2}{c}{(MHz)} & \colhead{(MHz)}       & \multicolumn{2}{c}{Area}         & \multicolumn{4}{c}{(mJy beam$^{-1}$)} & \multicolumn{2}{c}{Area}         & \multicolumn{4}{c}{(mJy)} \\
\colhead{}       & \colhead{(hh:mm:ss)} & \colhead{(dd:mm:ss)} & \colhead{}      & \colhead{}                    & \multicolumn{2}{c}{}      & \colhead{}            & \multicolumn{2}{c}{(arcmin$^2$)} & \multicolumn{4}{c}{}                  & \multicolumn{2}{c}{(arcmin$^2$)} & \multicolumn{4}{c}{} 
}
\decimals
\startdata
G233.753$-$00.193 & 07:30:02.2 & $-18$:32:43.2 & ch1 & B & 7183.3 & 4000 & 0.38 & 67.82 & 0.70 & 12.97 & 686.40 & 4.48  \\
 & & & & & 4897.5 & 256 & 1.10 & 216.76 & 4.84 & 14.62 & 1140.07 & 18.67  \\
 & & & & & 5153.5 & 256 & 0.94 & 180.97 & 4.70 & 12.82 & 1028.53 & 18.30  \\
 & & & & & 5409.5 & 256 & 0.92 & 171.71 & 4.03 & 13.06 & 985.59 & 16.04  \\
 & & & & & 5665.5 & 256 & 0.78 & 142.49 & 3.28 & 12.31 & 907.25 & 13.76  \\
 & & & & & 5921.6 & 256 & 0.74 & 129.71 & 3.71 & 10.71 & 829.21 & 14.89  \\
 & & & & & 6177.6 & 256 & 0.63 & 107.13 & 3.53 & 9.49 & 745.61 & 14.29  \\
 & & & & & 6433.6 & 256 & 0.62 & 100.88 & 6.08 & 4.96 & 489.40 & 17.59  \\
 & & & & & 7676.7 & 256 & 0.40 & 65.58 & 2.66 & 5.30 & 429.08 & 9.98  \\
 & & & & & 7932.7 & 256 & 0.39 & 61.27 & 3.46 & 4.10 & 363.42 & 11.53  \\
 & & & & & 8188.8 & 256 & 0.37 & 58.28 & 3.01 & 4.29 & 366.29 & 10.48  \\
 & & & & & 8444.8 & 256 & 0.38 & 57.22 & 3.35 & 4.00 & 347.99 & 11.15  \\
 & & & & & 8700.8 & 256 & 0.33 & 48.28 & 2.52 & 4.13 & 336.31 & 9.12  \\
 & & & & & 8956.8 & 256 & 0.35 & 47.79 & 3.04 & 3.62 & 306.01 & 10.00  \\
\enddata
\tablecomments{This table is available in its entirety in a machine-readable form in the online journal. A portion is shown here for guidance regarding its form and content.}
\tablenotetext{a}{Continuum detection quality factor (see text)}
\end{deluxetable*}
\end{longrotatetable}

\begin{longrotatetable}
\begin{deluxetable*}{lccccDcDD@{$\,\pm\,$}DDD@{$\,\pm\,$}D}
\centering
\tablewidth{0pt}
\tabletypesize{\scriptsize}
\tablecaption{Smoothed Continuum Detections\label{tab:continuum}}
\tablehead{
\colhead{Source} & \colhead{RA}         & \colhead{Dec.}       & \colhead{Field} & \colhead{QF\tablenotemark{a}} & \multicolumn{2}{c}{$\nu$} & \colhead{$\Delta\nu$} & \multicolumn{2}{c}{Beam}         & \multicolumn{4}{c}{$S_{C,P}$}         & \multicolumn{2}{c}{Region}       & \multicolumn{4}{c}{$S_{C,T}$} \\
\colhead{}       & \colhead{J2000}      & \colhead{J2000}      & \colhead{}      & \colhead{}                    & \multicolumn{2}{c}{(MHz)} & \colhead{(MHz)}       & \multicolumn{2}{c}{Area}         & \multicolumn{4}{c}{(mJy beam$^{-1}$)} & \multicolumn{2}{c}{Area}         & \multicolumn{4}{c}{(mJy)} \\
\colhead{}       & \colhead{(hh:mm:ss)} & \colhead{(dd:mm:ss)} & \colhead{}      & \colhead{}                    & \multicolumn{2}{c}{}      & \colhead{}            & \multicolumn{2}{c}{(arcmin$^2$)} & \multicolumn{4}{c}{}                  & \multicolumn{2}{c}{(arcmin$^2$)} & \multicolumn{4}{c}{} 
}
\decimals
\startdata
G233.753$-$00.193 & 07:30:02.5 & $-18$:32:35.2 & ch1 & A & 7183.3 & 4000 & 1.93 & 220.63 & 2.42 & 16.07 & 664.10 & 8.17  \\
 & & & & & 4897.5 & 256 & 1.93 & 323.85 & 7.21 & 16.30 & 1135.67 & 22.56  \\
 & & & & & 5153.5 & 256 & 1.93 & 307.24 & 8.04 & 14.55 & 1018.01 & 23.73  \\
 & & & & & 5409.5 & 256 & 1.93 & 294.36 & 6.85 & 15.01 & 977.50 & 20.75  \\
 & & & & & 5665.5 & 256 & 1.93 & 280.84 & 6.29 & 14.46 & 895.09 & 18.78  \\
 & & & & & 5921.6 & 256 & 1.93 & 270.68 & 7.47 & 12.82 & 818.87 & 20.92  \\
 & & & & & 6177.6 & 256 & 1.93 & 253.83 & 8.17 & 11.54 & 732.00 & 21.65  \\
 & & & & & 6433.6 & 256 & 1.93 & 238.27 & 12.95 & 7.08 & 507.68 & 26.23  \\
 & & & & & 7676.7 & 256 & 1.93 & 198.04 & 8.43 & 7.56 & 420.34 & 18.12  \\
 & & & & & 7932.7 & 256 & 1.93 & 190.27 & 11.28 & 5.82 & 350.02 & 21.08  \\
 & & & & & 8188.8 & 256 & 1.93 & 187.30 & 9.88 & 6.13 & 349.67 & 19.09  \\
 & & & & & 8444.8 & 256 & 1.93 & 182.55 & 11.07 & 5.60 & 328.90 & 20.41  \\
 & & & & & 8700.8 & 256 & 1.93 & 172.65 & 9.06 & 5.93 & 315.09 & 17.31  \\
 & & & & & 8956.8 & 256 & 1.93 & 168.60 & 10.22 & 5.23 & 284.94 & 18.27  \\
\enddata
\tablecomments{This table is available in its entirety in a machine-readable form in the online journal. A portion is shown here for guidance regarding its form and content.}
\tablenotetext{a}{Continuum detection quality factor (see text)}
\end{deluxetable*}
\end{longrotatetable}

\begin{longrotatetable}
\begin{deluxetable*}{lccDcD@{$\,\pm\,$}DD@{$\,\pm\,$}DD@{$\,\pm\,$}DD@{$\,\pm\,$}DD}
\centering
\tablewidth{0pt}
\tabletypesize{\scriptsize}
\tablecaption{Un-smoothed Peak RRL Detections\label{tab:unsmoothed_peak_line}}
\tablehead{
\colhead{Source} & \colhead{Field} & \colhead{RRL(s)\tablenotemark{a}} & \multicolumn{2}{c}{$\nu$\tablenotemark{b}} & \colhead{Comp.\tablenotemark{c}} & \multicolumn{4}{c}{$S_L$}             & \multicolumn{4}{c}{$V_{\rm LSR}$} & \multicolumn{4}{c}{$\Delta V$}    & \multicolumn{4}{c}{$S_{C}$}           & \multicolumn{2}{c}{SNR} \\
\colhead{}       & \colhead{}      & \colhead{}                        & \multicolumn{2}{c}{(MHz)}                  & \colhead{}                       & \multicolumn{4}{c}{(mJy beam$^{-1}$)} & \multicolumn{4}{c}{(km s$^{-1}$)} & \multicolumn{4}{c}{(km s$^{-1}$)} & \multicolumn{4}{c}{(mJy beam$^{-1}$)} & \multicolumn{2}{c}{}  
}
\decimals
\startdata
G233.753$-$00.193 & ch1 & H88$-$H112 & 6236.6 & a & 7.58 & 0.64 & 36.50 & 1.10 & 27.60 & 2.69 & 145.03 & 1.51 & 11.7  \\
 & & H100$-$H112 & 5414.3 & a & 7.94 & 0.82 & 36.70 & 1.30 & 26.60 & 3.17 & 176.06 & 1.89 & 9.6  \\
 & & H88$-$H95 & 8788.8 & a & 5.46 & 0.68 & 34.90 & 2.20 & 36.31 & 5.28 & 48.95 & 1.84 & 7.9  \\
 & & H106$-$H112 & 5010.3 & a & 8.33 & 1.05 & 36.90 & 1.90 & 31.33 & 4.56 & 206.36 & 2.63 & 7.8  \\
 & & H100$-$H105 & 5914.1 & a & 7.74 & 1.34 & 37.50 & 1.80 & 21.46 & 4.31 & 138.60 & 2.78 & 5.7  \\
 & & H112 & 4618.8 & a & \nodata & \nodata & \nodata & \nodata & \nodata & \nodata & 240.98 & 7.32 & \nodata  \\
 & & H111 & 4744.2 & a & \nodata & \nodata & \nodata & \nodata & \nodata & \nodata & 223.74 & 6.61 & \nodata  \\
 & & H110 & 4874.2 & a & \nodata & \nodata & \nodata & \nodata & \nodata & \nodata & 225.37 & 6.95 & \nodata  \\
 & & H109 & 5008.9 & a & \nodata & \nodata & \nodata & \nodata & \nodata & \nodata & 208.47 & 7.34 & \nodata  \\
 & & H108 & 5148.7 & a & \nodata & \nodata & \nodata & \nodata & \nodata & \nodata & 192.22 & 9.50 & \nodata  \\
 & & H107 & 5293.7 & a & \nodata & \nodata & \nodata & \nodata & \nodata & \nodata & 182.74 & 6.84 & \nodata  \\
 & & H106 & 5444.3 & a & \nodata & \nodata & \nodata & \nodata & \nodata & \nodata & 169.35 & 5.77 & \nodata  \\
 & & H105 & 5600.6 & a & \nodata & \nodata & \nodata & \nodata & \nodata & \nodata & 160.07 & 5.69 & \nodata  \\
 & & H104 & 5762.9 & a & \nodata & \nodata & \nodata & \nodata & \nodata & \nodata & 146.62 & 5.87 & \nodata  \\
 & & H103 & 5931.5 & a & 9.74 & 2.69 & 39.30 & 3.00 & 21.08 & 8.97 & 134.30 & 5.14 & 3.9  \\
 & & H102 & 6106.9 & a & \nodata & \nodata & \nodata & \nodata & \nodata & \nodata & 127.23 & 5.52 & \nodata  \\
 & & H100 & 6478.8 & a & \nodata & \nodata & \nodata & \nodata & \nodata & \nodata & 108.69 & 5.81 & \nodata  \\
 & & H94 & 7792.9 & a & \nodata & \nodata & \nodata & \nodata & \nodata & \nodata & 39.89 & 4.76 & \nodata  \\
 & & H92 & 8309.4 & a & \nodata & \nodata & \nodata & \nodata & \nodata & \nodata & 57.22 & 4.79 & \nodata  \\
 & & H91 & 8584.8 & a & \nodata & \nodata & \nodata & \nodata & \nodata & \nodata & 56.35 & 3.94 & \nodata  \\
 & & H89 & 9173.3 & a & \nodata & \nodata & \nodata & \nodata & \nodata & \nodata & 46.92 & 4.00 & \nodata  \\
 & & H88 & 9487.8 & a & 7.17 & 1.48 & 35.40 & 3.40 & 31.75 & 11.24 & 42.27 & 3.37 & 5.3  \\
\enddata
\tablecomments{This table is available in its entirety in a machine-readable form in the online journal. A portion is shown here for guidance regarding its form and content.}
\tablenotetext{a}{RRL ranges indicate average RRL spectra, but only of those RRL transitions within that range that are in this table (i.e., H90 is not included in the H88$-$H112 average).}
\tablenotetext{b}{RRL rest frequency. For average RRL spectra, the weighted average RRL rest frequency.}
\tablenotetext{c}{``a'' is the brightest Gaussian component, ``b'' is the second brightest, etc.}
\end{deluxetable*}
\end{longrotatetable}

\begin{longrotatetable}
\begin{deluxetable*}{lccDcD@{$\,\pm\,$}DD@{$\,\pm\,$}DD@{$\,\pm\,$}DD@{$\,\pm\,$}DD}
\centering
\tablewidth{0pt}
\tabletypesize{\scriptsize}
\tablecaption{Smoothed Peak RRL Detections\label{tab:smoothed_peak_line}}
\tablehead{
\colhead{Source} & \colhead{Field} & \colhead{RRL(s)\tablenotemark{a}} & \multicolumn{2}{c}{$\nu$\tablenotemark{b}} & \colhead{Comp.\tablenotemark{c}} & \multicolumn{4}{c}{$S_L$}             & \multicolumn{4}{c}{$V_{\rm LSR}$} & \multicolumn{4}{c}{$\Delta V$}    & \multicolumn{4}{c}{$S_{C}$}           & \multicolumn{2}{c}{SNR} \\
\colhead{}       & \colhead{}      & \colhead{}                        & \multicolumn{2}{c}{(MHz)}                  & \colhead{}                       & \multicolumn{4}{c}{(mJy beam$^{-1}$)} & \multicolumn{4}{c}{(km s$^{-1}$)} & \multicolumn{4}{c}{(km s$^{-1}$)} & \multicolumn{4}{c}{(mJy beam$^{-1}$)} & \multicolumn{2}{c}{}  
}
\decimals
\startdata
G233.753$-$00.193 & ch1 & H88$-$H112 & 6326.4 & a & 16.70 & 0.82 & 35.80 & 0.60 & 24.66 & 1.40 & 279.29 & 1.83 & 20.1  \\
 & & H100$-$H112 & 5440.4 & a & 16.42 & 0.99 & 36.10 & 0.70 & 24.94 & 1.74 & 315.59 & 2.22 & 16.3  \\
 & & H88$-$H95 & 8852.0 & a & 17.51 & 1.47 & 35.10 & 1.00 & 24.11 & 2.34 & 175.60 & 3.23 & 11.8  \\
 & & H106$-$H112 & 4992.8 & a & 16.44 & 1.30 & 36.60 & 1.10 & 29.05 & 2.66 & 337.54 & 3.15 & 12.5  \\
 & & H100$-$H105 & 5916.6 & a & 16.33 & 1.55 & 35.60 & 1.00 & 22.29 & 2.44 & 292.31 & 3.27 & 10.4  \\
 & & H112 & 4618.8 & a & \nodata & \nodata & \nodata & \nodata & \nodata & \nodata & 357.49 & 8.27 & \nodata  \\
 & & H111 & 4744.2 & a & \nodata & \nodata & \nodata & \nodata & \nodata & \nodata & 341.78 & 7.40 & \nodata  \\
 & & H110 & 4874.2 & a & 21.38 & 4.45 & 33.70 & 1.70 & 16.66 & 4.45 & 349.38 & 7.91 & 4.9  \\
 & & H109 & 5008.9 & a & \nodata & \nodata & \nodata & \nodata & \nodata & \nodata & 340.55 & 8.20 & \nodata  \\
 & & H108 & 5148.7 & a & \nodata & \nodata & \nodata & \nodata & \nodata & \nodata & 335.95 & 12.10 & \nodata  \\
 & & H107 & 5293.7 & a & 19.97 & 3.47 & 30.60 & 2.10 & 24.10 & 5.31 & 320.83 & 7.43 & 5.8  \\
 & & H106 & 5444.3 & a & 19.87 & 3.35 & 35.40 & 1.80 & 22.19 & 4.40 & 317.49 & 7.03 & 5.9  \\
 & & H105 & 5600.6 & a & 15.54 & 3.46 & 36.50 & 2.70 & 23.62 & 7.74 & 306.62 & 7.09 & 4.7  \\
 & & H104 & 5762.9 & a & 14.68 & 3.15 & 35.20 & 3.20 & 28.49 & 10.10 & 302.97 & 6.89 & 5.0  \\
 & & H103 & 5931.5 & a & 19.62 & 3.41 & 38.20 & 2.00 & 23.44 & 5.29 & 291.27 & 7.17 & 5.9  \\
 & & H102 & 6106.9 & a & \nodata & \nodata & \nodata & \nodata & \nodata & \nodata & 283.81 & 5.99 & \nodata  \\
 & & H100 & 6478.8 & a & \nodata & \nodata & \nodata & \nodata & \nodata & \nodata & 257.65 & 7.02 & \nodata  \\
 & & H94 & 7792.9 & a & \nodata & \nodata & \nodata & \nodata & \nodata & \nodata & 132.06 & 11.06 & \nodata  \\
 & & H92 & 8309.4 & a & \nodata & \nodata & \nodata & \nodata & \nodata & \nodata & 197.15 & 7.15 & \nodata  \\
 & & H91 & 8584.8 & a & 18.67 & 2.60 & 34.40 & 1.80 & 26.26 & 4.48 & 180.10 & 5.86 & 7.2  \\
 & & H89 & 9173.3 & a & 15.90 & 3.11 & 35.60 & 2.00 & 21.27 & 5.30 & 173.84 & 6.24 & 5.2  \\
 & & H88 & 9487.8 & a & 16.17 & 2.53 & 35.80 & 2.20 & 28.62 & 5.70 & 165.91 & 5.89 & 6.5  \\
\enddata
\tablecomments{This table is available in its entirety in a machine-readable form in the online journal. A portion is shown here for guidance regarding its form and content.}
\tablenotetext{a}{RRL ranges indicate average RRL spectra, but only of those RRL transitions within that range that are in this table (i.e., H90 is not included in the H88$-$H112 average).}
\tablenotetext{b}{RRL rest frequency. For average RRL spectra, the weighted average RRL rest frequency.}
\tablenotetext{c}{``a'' is the brightest Gaussian component, ``b'' is the second brightest, etc.}
\end{deluxetable*}
\end{longrotatetable}

\begin{longrotatetable}
\begin{deluxetable*}{lccDcD@{$\,\pm\,$}DD@{$\,\pm\,$}DD@{$\,\pm\,$}DD@{$\,\pm\,$}DD}
\centering
\tablewidth{0pt}
\tabletypesize{\scriptsize}
\tablecaption{Un-smoothed Total RRL Detections\label{tab:unsmoothed_total_line}}
\tablehead{
\colhead{Source} & \colhead{Field} & \colhead{RRL(s)\tablenotemark{a}} & \multicolumn{2}{c}{$\nu$\tablenotemark{b}} & \colhead{Comp.\tablenotemark{c}} & \multicolumn{4}{c}{$S_L$}             & \multicolumn{4}{c}{$V_{\rm LSR}$} & \multicolumn{4}{c}{$\Delta V$}    & \multicolumn{4}{c}{$S_{C}$}           & \multicolumn{2}{c}{SNR} \\
\colhead{}       & \colhead{}      & \colhead{}                        & \multicolumn{2}{c}{(MHz)}                  & \colhead{}                       & \multicolumn{4}{c}{(mJy)} & \multicolumn{4}{c}{(km s$^{-1}$)} & \multicolumn{4}{c}{(km s$^{-1}$)} & \multicolumn{4}{c}{(mJy)} & \multicolumn{2}{c}{}  
}
\decimals
\startdata
G233.753$-$00.193 & ch1 & H88$-$H112 & 6520.6 & a & 38.38 & 1.77 & 35.80 & 0.50 & 23.05 & 1.25 & 610.85 & 3.80 & 21.5  \\
 & & H100$-$H112 & 5467.7 & a & 43.53 & 2.24 & 35.80 & 0.60 & 23.11 & 1.38 & 766.18 & 4.81 & 19.2  \\
 & & H88$-$H95 & 8781.9 & a & 27.34 & 2.40 & 35.80 & 1.00 & 23.07 & 2.35 & 277.60 & 5.16 & 11.3  \\
 & & H106$-$H112 & 4997.9 & a & 44.86 & 3.05 & 36.60 & 0.80 & 24.28 & 1.91 & 855.47 & 6.74 & 14.5  \\
 & & H100$-$H105 & 5936.0 & a & 42.73 & 3.14 & 35.10 & 0.80 & 22.48 & 1.91 & 676.80 & 6.68 & 13.4  \\
 & & H112 & 4618.8 & a & 40.98 & 8.56 & 37.00 & 3.60 & 31.14 & 13.12 & 1025.05 & 18.64 & 5.4  \\
 & & H111 & 4744.2 & a & 35.15 & 6.87 & 36.70 & 3.00 & 31.19 & 7.90 & 914.77 & 16.65 & 5.2  \\
 & & H110 & 4874.2 & a & 49.17 & 7.75 & 36.70 & 2.00 & 25.96 & 5.72 & 907.94 & 16.81 & 6.6  \\
 & & H109 & 5008.9 & a & 46.38 & 6.70 & 36.00 & 1.50 & 21.34 & 3.66 & 772.29 & 13.74 & 6.9  \\
 & & H108 & 5148.7 & a & \nodata & \nodata & \nodata & \nodata & \nodata & \nodata & 750.19 & 22.26 & \nodata  \\
 & & H107 & 5293.7 & a & 40.97 & 6.88 & 31.40 & 2.10 & 25.41 & 5.15 & 788.86 & 15.33 & 6.0  \\
 & & H106 & 5444.3 & a & 53.37 & 7.25 & 37.50 & 1.70 & 24.86 & 3.98 & 806.48 & 16.10 & 7.3  \\
 & & H105 & 5600.6 & a & 42.98 & 7.05 & 35.60 & 1.90 & 23.21 & 4.62 & 749.93 & 14.99 & 6.1  \\
 & & H104 & 5762.9 & a & 43.62 & 7.18 & 33.90 & 2.10 & 25.66 & 5.57 & 740.65 & 15.72 & 6.2  \\
 & & H103 & 5931.5 & a & 50.28 & 7.56 & 36.40 & 1.60 & 22.34 & 4.04 & 682.28 & 15.82 & 6.7  \\
 & & H102 & 6106.9 & a & 38.45 & 6.40 & 33.90 & 1.90 & 23.03 & 5.49 & 637.72 & 13.01 & 6.3  \\
 & & H100 & 6478.8 & a & 35.21 & 6.83 & 34.00 & 1.70 & 17.50 & 4.23 & 498.37 & 12.52 & 5.2  \\
 & & H94 & 7792.9 & a & \nodata & \nodata & \nodata & \nodata & \nodata & \nodata & 85.94 & 8.05 & \nodata  \\
 & & H92 & 8309.4 & a & 31.54 & 5.89 & 37.90 & 3.30 & 30.00 & 12.15 & 341.54 & 12.46 & 6.1  \\
 & & H91 & 8584.8 & a & 31.11 & 4.40 & 34.80 & 1.60 & 23.84 & 3.96 & 291.04 & 9.57 & 7.0  \\
 & & H89 & 9173.3 & a & 25.25 & 4.60 & 36.10 & 2.30 & 25.11 & 5.95 & 293.13 & 10.00 & 5.6  \\
 & & H88 & 9487.8 & a & 29.44 & 4.65 & 36.00 & 1.90 & 24.71 & 4.89 & 289.59 & 10.12 & 6.4  \\
\enddata
\tablecomments{This table is available in its entirety in a machine-readable form in the online journal. A portion is shown here for guidance regarding its form and content.}
\tablenotetext{a}{RRL ranges indicate average RRL spectra, but only of those RRL transitions within that range that are in this table (i.e., H90 is not included in the H88$-$H112 average).}
\tablenotetext{b}{RRL rest frequency. For average RRL spectra, the weighted average RRL rest frequency.}
\tablenotetext{c}{``a'' is the brightest Gaussian component, ``b'' is the second brightest, etc.}
\end{deluxetable*}
\end{longrotatetable}

\begin{longrotatetable}
\begin{deluxetable*}{lccDcD@{$\,\pm\,$}DD@{$\,\pm\,$}DD@{$\,\pm\,$}DD@{$\,\pm\,$}DD}
\centering
\tablewidth{0pt}
\tabletypesize{\scriptsize}
\tablecaption{Smoothed Total RRL Detections\label{tab:smoothed_total_line}}
\tablehead{
\colhead{Source} & \colhead{Field} & \colhead{RRL(s)\tablenotemark{a}} & \multicolumn{2}{c}{$\nu$\tablenotemark{b}} & \colhead{Comp.\tablenotemark{c}} & \multicolumn{4}{c}{$S_L$}             & \multicolumn{4}{c}{$V_{\rm LSR}$} & \multicolumn{4}{c}{$\Delta V$}    & \multicolumn{4}{c}{$S_{C}$}           & \multicolumn{2}{c}{SNR} \\
\colhead{}       & \colhead{}      & \colhead{}                        & \multicolumn{2}{c}{(MHz)}                  & \colhead{}                       & \multicolumn{4}{c}{(mJy)} & \multicolumn{4}{c}{(km s$^{-1}$)} & \multicolumn{4}{c}{(km s$^{-1}$)} & \multicolumn{4}{c}{(mJy)} & \multicolumn{2}{c}{}  
}
\decimals
\startdata
G233.753$-$00.193 & ch1 & H88$-$H112 & 6556.6 & a & 37.07 & 1.68 & 35.90 & 0.50 & 23.29 & 1.22 & 581.32 & 3.64 & 21.8  \\
 & & H100$-$H112 & 5469.2 & a & 43.41 & 2.24 & 36.00 & 0.60 & 23.20 & 1.38 & 756.18 & 4.82 & 19.2  \\
 & & H88$-$H95 & 8708.6 & a & 24.07 & 2.18 & 35.40 & 1.00 & 22.76 & 2.38 & 235.39 & 4.66 & 10.9  \\
 & & H106$-$H112 & 5008.2 & a & 44.24 & 3.09 & 36.80 & 0.80 & 24.62 & 1.99 & 843.35 & 6.87 & 14.1  \\
 & & H100$-$H105 & 5929.4 & a & 43.13 & 3.07 & 35.20 & 0.80 & 22.47 & 1.85 & 669.18 & 6.52 & 13.9  \\
 & & H112 & 4618.8 & a & 43.16 & 9.07 & 36.90 & 3.10 & 27.81 & 10.17 & 1026.64 & 19.31 & 5.2  \\
 & & H111 & 4744.2 & a & 34.03 & 6.71 & 37.50 & 3.30 & 33.76 & 9.13 & 904.07 & 16.67 & 5.3  \\
 & & H110 & 4874.2 & a & 49.58 & 7.80 & 37.30 & 2.10 & 26.28 & 5.83 & 901.59 & 16.98 & 6.6  \\
 & & H109 & 5008.9 & a & 45.31 & 6.57 & 35.70 & 1.50 & 20.81 & 3.61 & 758.60 & 13.29 & 6.9  \\
 & & H108 & 5148.7 & a & \nodata & \nodata & \nodata & \nodata & \nodata & \nodata & 747.02 & 21.82 & \nodata  \\
 & & H107 & 5293.7 & a & 41.17 & 6.52 & 32.80 & 2.30 & 29.89 & 5.69 & 761.96 & 15.79 & 6.3  \\
 & & H106 & 5444.3 & a & 53.14 & 7.06 & 38.40 & 1.70 & 26.14 & 4.10 & 790.11 & 16.07 & 7.5  \\
 & & H105 & 5600.6 & a & 43.02 & 7.23 & 35.50 & 1.90 & 23.23 & 5.04 & 736.37 & 15.15 & 6.1  \\
 & & H104 & 5762.9 & a & 42.86 & 7.03 & 34.10 & 2.10 & 25.67 & 5.36 & 726.10 & 15.54 & 6.2  \\
 & & H103 & 5931.5 & a & 51.45 & 7.72 & 36.30 & 1.70 & 22.59 & 4.09 & 672.61 & 16.23 & 6.7  \\
 & & H102 & 6106.9 & a & 39.11 & 6.39 & 34.40 & 1.70 & 21.49 & 4.34 & 631.17 & 13.00 & 6.2  \\
 & & H100 & 6478.8 & a & 34.89 & 6.46 & 34.80 & 1.80 & 19.34 & 4.51 & 498.95 & 12.42 & 5.5  \\
 & & H94 & 7792.9 & a & \nodata & \nodata & \nodata & \nodata & \nodata & \nodata & 64.81 & 5.29 & \nodata  \\
 & & H92 & 8309.4 & a & 30.31 & 5.52 & 36.70 & 2.60 & 26.54 & 8.76 & 304.65 & 11.36 & 6.1  \\
 & & H91 & 8584.8 & a & 28.89 & 4.25 & 35.40 & 1.80 & 25.35 & 4.61 & 264.89 & 9.41 & 6.8  \\
 & & H89 & 9173.3 & a & 25.82 & 4.76 & 35.30 & 1.90 & 21.20 & 4.98 & 270.30 & 9.54 & 5.5  \\
 & & H88 & 9487.8 & a & 26.79 & 4.84 & 34.50 & 1.90 & 21.98 & 4.87 & 256.96 & 10.00 & 5.6  \\
\enddata
\tablecomments{This table is available in its entirety in a machine-readable form in the online journal. A portion is shown here for guidance regarding its form and content.}
\tablenotetext{a}{RRL ranges indicate average RRL spectra, but only of those RRL transitions within that range that are in this table (i.e., H90 is not included in the H88$-$H112 average).}
\tablenotetext{b}{RRL rest frequency. For average RRL spectra, the weighted average RRL rest frequency.}
\tablenotetext{c}{``a'' is the brightest Gaussian component, ``b'' is the second brightest, etc.}
\end{deluxetable*}
\end{longrotatetable}

\begin{figure*}
  \centering
  \includegraphics[width=\textwidth]{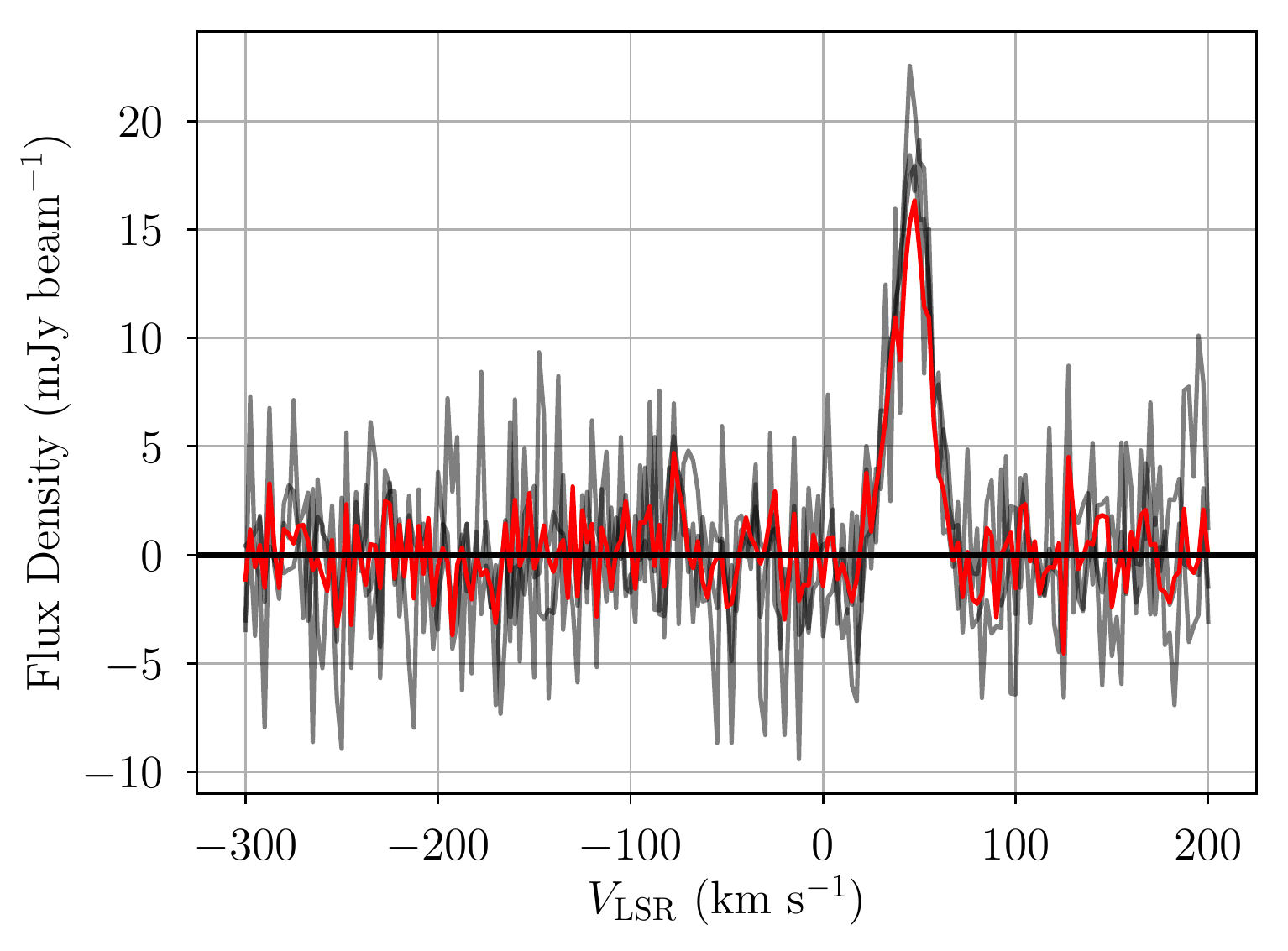}
  \caption{Example SHRDS spectrum showing the spectral sensitivity
    improvement in the mosaic data. The gray curves are
    \(\langle\text{Hn}\alpha\rangle\) spectra toward G293.994$-$00.934
    in three different SHRDS fields. The red curve is the
    \(\langle\text{Hn}\alpha\rangle\) spectrum toward the same source
    in the mosaic ``mos083''. Each spectrum is extracted from the
    brightest continuum pixel in the smoothed, 4\ghz\ bandwidth
    continuum images. The spectral rms is \(0.84\), \(1.91\), and
    \(1.35\,\text{mJy beam\(^{-1}\)}\) in fields ``shrds219'',
    ``g293.936$-$'', and ``g293.994$-$,'' respectively. In the mosaic
    spectrum, the spectral rms is \(0.75\,\text{mJy beam\(^{-1}\)}\).}
  \label{fig:spectrum}
\end{figure*}

\begin{figure*}
  \centering
  \includegraphics[width=\textwidth]{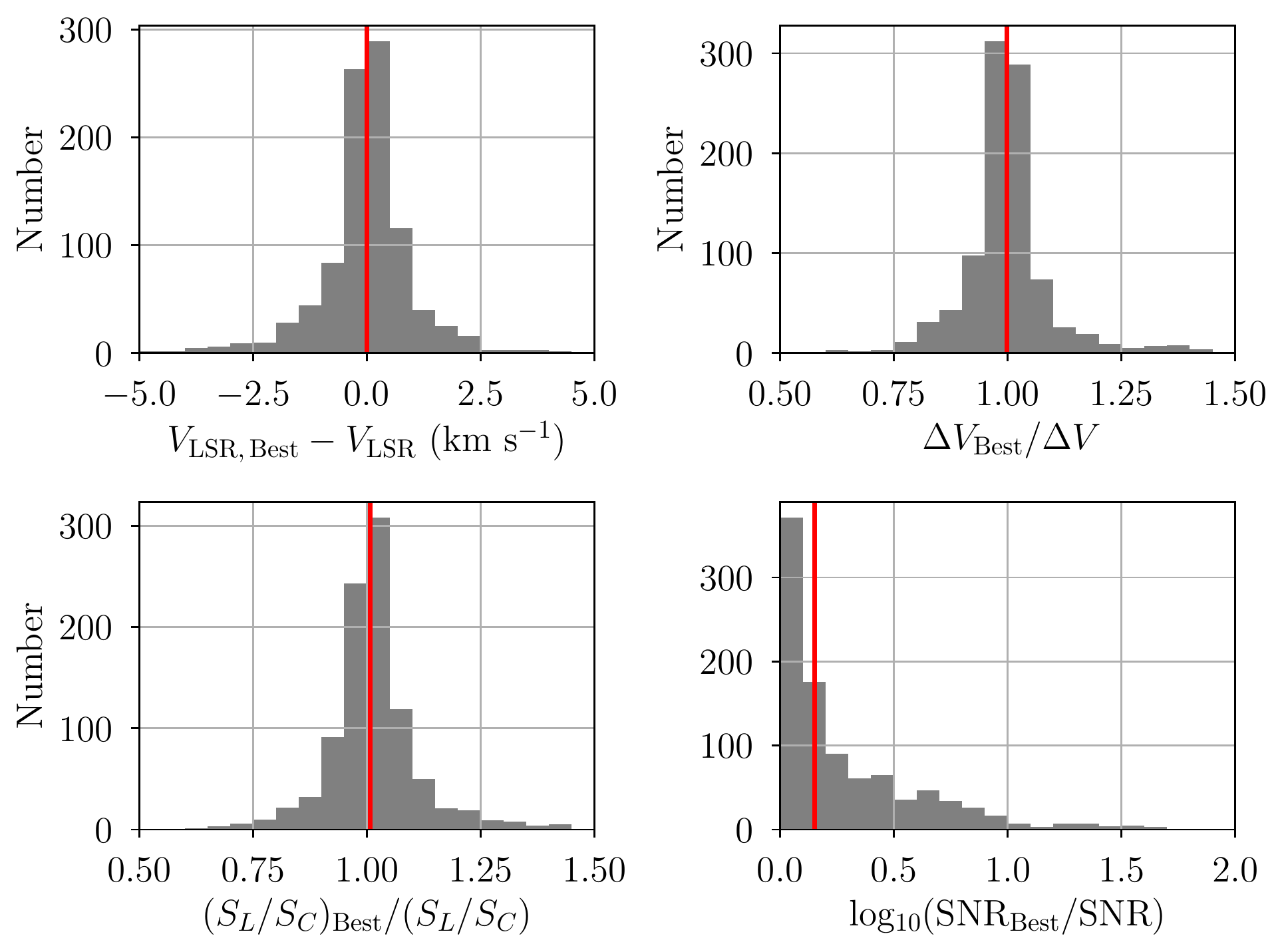}
  \caption{A comparison of the RRL parameters for sources with
    detections in multiple fields. The top-left panel is the
    difference between the fitted LSR velocity for the detection with
    the highest RRL SNR and each other detection of the same source,
    the top-right panel is the ratio of the fitted FWHM line width,
    the bottom-left panel is the ratio of the RRL-to-continuum ratios,
    and the bottom right panel is the ratio of RRL SNR. For nebulae
    with multiple RRL velocity components, only the brightest RRL
    component is compared. The median LSR velocity difference is
    \(0.0\kms\), the median FWHM line width ratio and RRL-to-continuum
    are both \(1.0\), and the median RRL SNR ratio is \(1.4\). These
    median values are indicated by the vertical red line in each
    panel. Our data analysis techniques recover the same RRL
    properties for sources detected independently in different
    fields.}
  \label{fig:multiple}
\end{figure*}

\subsection{Radio Recombination Line Catalog}

The RRL properties of the SHRDS nebulae are summarized in four
different catalogs. The RRL parameters of spectra extracted from the
brightest continuum pixel are given in
Table~\ref{tab:unsmoothed_peak_line} (non-smoothed) and
Table~\ref{tab:smoothed_peak_line} (smoothed). The RRL parameters
derived from the watershed region spectra are given in
Table~\ref{tab:unsmoothed_total_line} (non-smoothed) and
Table~\ref{tab:smoothed_total_line} (smoothed). For each source, we
list the \textit{WISE} Catalog name and field name.  Then, for each
RRL transition and \(\langle\text{Hn}\alpha\rangle\) spectrum, we list
the RRL transition or range of averaged RRLs, RRL rest frequency or
weighted-average RRL rest frequency, \(\nu\), the velocity component
identifier, fitted Gaussian peak RRL brightness or flux density,
\(S_L\), fitted center LSR velocity, \(V_{\rm LSR}\), fitted
full-width at half-maximum (FWHM) line width, \(\Delta V\), continuum
brightness or flux density, \(S_C\), and signal-to-noise ratio,
SNR. The average RRL ranges only include those transitions for which
there are individual transitions listed in the tables (for example,
H90 is not observed and not listed in the table, therefore the
``H88-H112'' average RRL does not include H90). The average rest
frequencies are weighted using the same weights as the
\(\langle\text{Hn}\alpha\rangle\) spectra (see Equation~4 in
\citet{wenger2019b}). We fit multiple Gaussian components in some
cases, and the velocity component identifier is ``a'' for the
brightest component, ``b'' for the second brightest, etc. The
continuum brightness or flux density is the median value of the
line-free channels, and the uncertainty is the spectral rms.  The SNR
is estimated following \citet{lenz1992},
\begin{equation}
  SNR = 0.7\left(\frac{S_L}{\rm rms}\right)\left(\frac{\Delta
    V}{\Delta v}\right)^{0.5},
\end{equation}
where rms is the spectral rms and \(\Delta v\) is the channel width.

To test the reproducibility of our RRL catalog, we use the
\(\langle\text{Hn}\alpha\rangle\) RRL parameters of nebulae with
multiple detections. A \textit{WISE} Catalog source can appear in the
RRL tables several times if it is detected in multiple fields and/or
mosaics. For example, Figure~\ref{fig:spectrum} shows four
\(\langle\text{Hn}\alpha\rangle\) spectra toward
G293.994$-$00.934. The spectrum extracted from the mosaic data has the
best sensitivity, as expected. For each source, we identify the RRL
detection with the highest SNR (the ``best'' detection).
Figure~\ref{fig:multiple} shows distributions of the differences
between the best detections and every other detection for various RRL
parameters. Shown are the difference between the LSR velocities, the
ratio of FWHM linewidths, the ratio of RRL-to-continuum brightness
ratios, and the ratio of RRL SNRs. We only include the brightest RRL
component if the source has multiple velocity components. For an
optically thin nebula in local thermodynamic equilibrium (LTE), the
RRL-to-continuum ratio, \(S_L/S_C\), scales as \(S_L/S_C \propto
\nu^{1.1}\), where \(\nu\) is the RRL rest frequency
\citep{wenger2019b}. We scale each RRL-to-continuum brightness ratio
to 9\ghz. The median velocity difference is \(0.0\kms\), and the
median ratio of line widths and RRL-to-continuum ratios are both
\(1.0\). We therefore conclude that our data reduction and analysis
procedures are recovering the same RRL properties for multiple
detections.

\section{Properties of SHRDS Nebulae}

The current census of known \hii\ regions in the \textit{WISE} Catalog
is comprised of 2376 nebulae. This population is the sum of
\hii\ regions discovered by the HRDS and SHRDS, together with those
previously known nebulae listed in the \textit{WISE} Catalog. The HRDS
and SHRDS together have discovered \({\sim}1400\) nebulae, more than
doubling the \textit{WISE} Catalog population of Milky Way
\hii\ regions. The SHRDS adds 436 nebulae to this census, a 130\%
increase in the range \(259^\circ < \ell < 346^\circ\). Here we
compare the properties of the SHRDS nebulae which the current
\hii\ region census.

\begin{figure*}
  \centering
  \includegraphics[height=0.8\textheight]{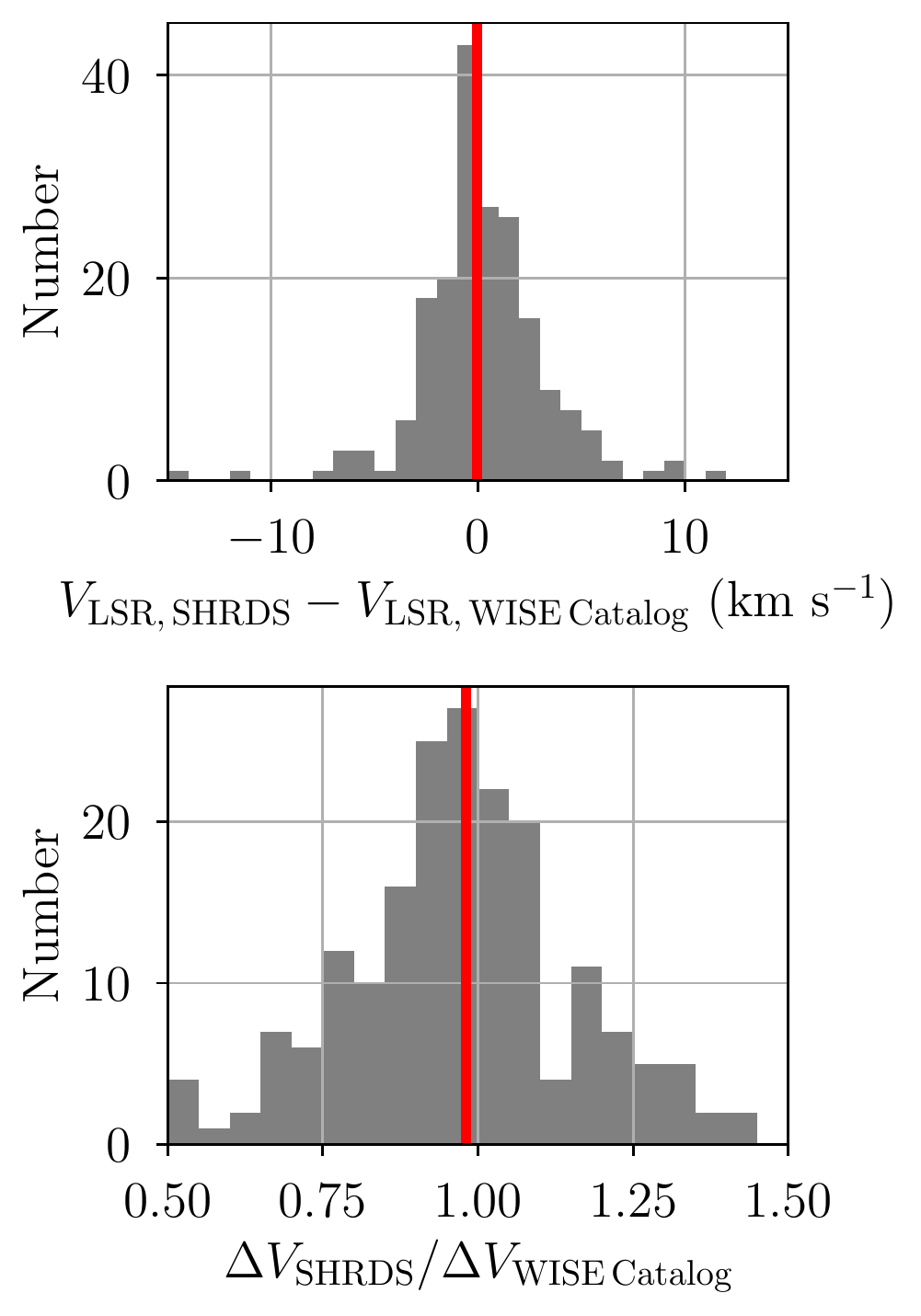}
  \caption{A comparison between the SHRDS RRL parameters and those
    previously measured for 207 \hii\ regions. The top panel is the
    distribution of the difference between the SHRDS LSR velocity,
    \(V_{\rm LSR}\), and the previously measured LSR velocity from the
    \textit{WISE} Catalog. The bottom panel is the distribution of the
    ratio of the SHRDS FWHM line width, \(\Delta V\), to the
    previously measured line width. The red lines indicate the median
    of each distribution: \(0.0\kms\) (top) and \(0.98\) (bottom). For
    nebulae with multiple RRL velocity components, only the brightest
    RRL component is compared.  We exclude nine nebulae with LSR
    velocity differences greater than \(10\kms\). The FWHM line width
    dispersion may stem from the ATCA probing a different volume of
    gas than the single dish observations.}
  \label{fig:previous_compare}
\end{figure*}

\begin{figure*}
  \centering
  \includegraphics[height=0.8\textheight]{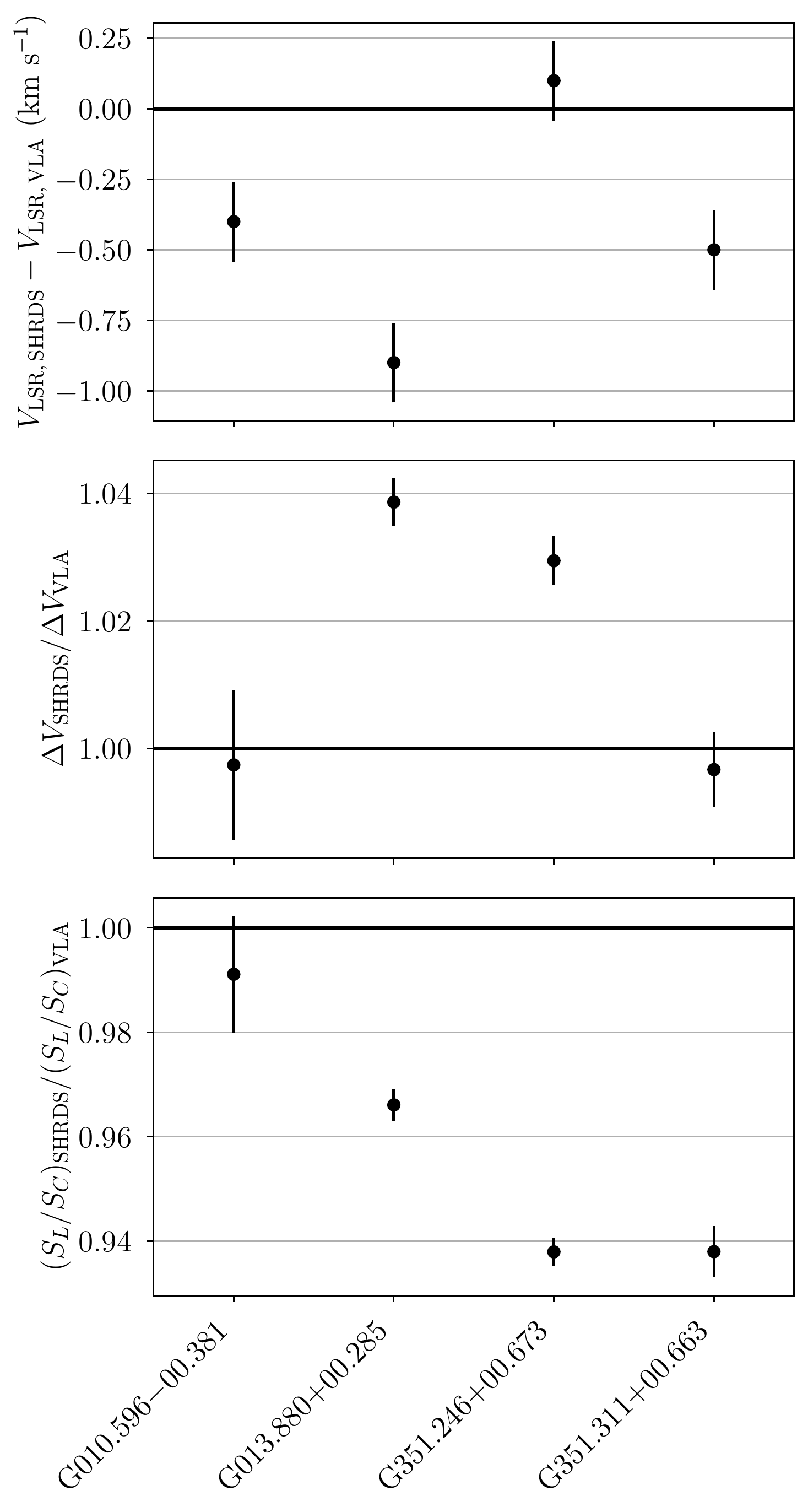}
  \caption{A comparison between the SHRDS RRL and radio continuum data
    and those previously measured by the VLA \citep{wenger2019b} for
    the four nebulae in common between the two surveys. The top panel
    is the difference between the SHRDS and VLA RRL LSR velocity,
    \(V_{\rm LSR}\).  The middle panel is the ratio of the SHRDS RRL
    FWHM line width, \(\Delta V\), to that measured by the VLA. The
    bottom panel is the ratio of the RRL-to-continuum brightness
    ratio, \(S_L/S_C\), measured by the SHRDS to that measured by the
    VLA. The error bars represent the \(1\sigma\) uncertainties, and
    the horizontal lines represent parity between the two datasets.
    There is a systematic discrepancy between the SHRDS and VLA
    RRL-to-continuum brightness ratios.}
  \label{fig:vla_compare}
\end{figure*}

\subsection{Previously Known Nebulae}

The SHRDS RRL catalog includes 206 detections toward previously known
\hii\ regions.  The total flux density measurements of these nebulae
will vary between experiments due to differences in observing
frequency, beam size, and the loss of flux in interferometric
observations. We can nonetheless use the SHRDS detections of
previously known \hii\ regions to assess the efficacy of the
experiment by comparing the kinematic RRL properties (i.e., center
velocities and line widths) and RRL-to-continuum brightness ratios.

We find that there are no systematic discrepancies between the SHRDS
RRL LSR velocities and FWHM line widths compared to previous single
dish measurements. A nebula's RRL LSR velocity and FWHM line width
should be independent of the telescope if each telescope is seeing the
same volume of emitting gas. Therefore, we expect to see some
differences between the SHRDS RRL parameters and those previously
measured by single dish telescopes because the ATCA probes a smaller
and more compact volume.  Figure~\ref{fig:previous_compare} compares
the LSR velocities and FWHM line widths from the SHRDS against
previous single dish measurements listed in the \textit{WISE}
Catalog. We use the SHRDS detection with the highest RRL SNR to make
this comparison. There are nine nebulae with LSR velocity differences
greater than \(10\kms\): seven from \citet{caswell1987}, one from
\citet{lockman1989}, and one from \citet{anderson2011}. These outliers
are likely due to misassociations with \textit{WISE} Catalog objects
due to the poor angular resolution of the previous studies or
confusion in the SHRDS. Excluding these outliers, the median
difference between the SHRDS and \textit{WISE} Catalog RRL LSR
velocities is \(0.0\kms\) with a standard deviation of \(2.7\kms\). In
contrast, the median FWHM line width ratio is \(0.98\) with a standard
deviation of \(0.23\). These distributions are consistent with no
systematic difference between the SHRDS and single dish studies. The
large dispersion of the FWHM line width distribution, however,
supports our expectation that the ATCA is probing a different volume
of gas.

Although we find no systematic differences between the SHRDS and
previous single dish measurements, we do find some discrepancies
compared to previous interferometric observations with the VLA. There
are four nebulae observed in both the SHRDS and the
\citet{wenger2019b} VLA survey: G010.596$-$00.381, G013.880+00.285,
G351.246+00.673, and G351.311+00.663. All four nebulae are robust
detections with a RRL SNR \(>100\) in both surveys. Overall they also
have excellent continuum data with QF A in both surveys, except for
G010.596$-$00.381 with QF C in the SHRDS. The differences between the
SHRDS and VLA RRL LSR velocities and FWHM line widths are small (see
top two panels of Figure~\ref{fig:vla_compare}). The median LSR
velocity difference and FWHM line width ratio are \(-0.5\kms\) and
\(1.01\), respectively. The bottom panel of
Figure~\ref{fig:vla_compare} shows the ratio of the SHRDS and VLA
RRL-to-continuum ratios after scaling each to a 8\ghz. Here there is a
systematic discrepency between the two surveys; the median SHRDS
RRL-to-continuum ratio is 95\% of the VLA ratio.  These differences
may be a consequence of the much larger synthesized beam size in the
SHRDS (\({\sim}100\arcsper\)) probing a larger volume of gas compared
to the VLA \({\sim}10\arcsper\), or they may be due to optical depth
or non-LTE effects. Since the RRL-to-continuum ratio is used to derive
electron temperatures and infer nebular metallicities in Galactic
\hii\ regions, this systematic offset must be investigated further.

\begin{figure*}
  \centering
  \includegraphics[width=\textwidth]{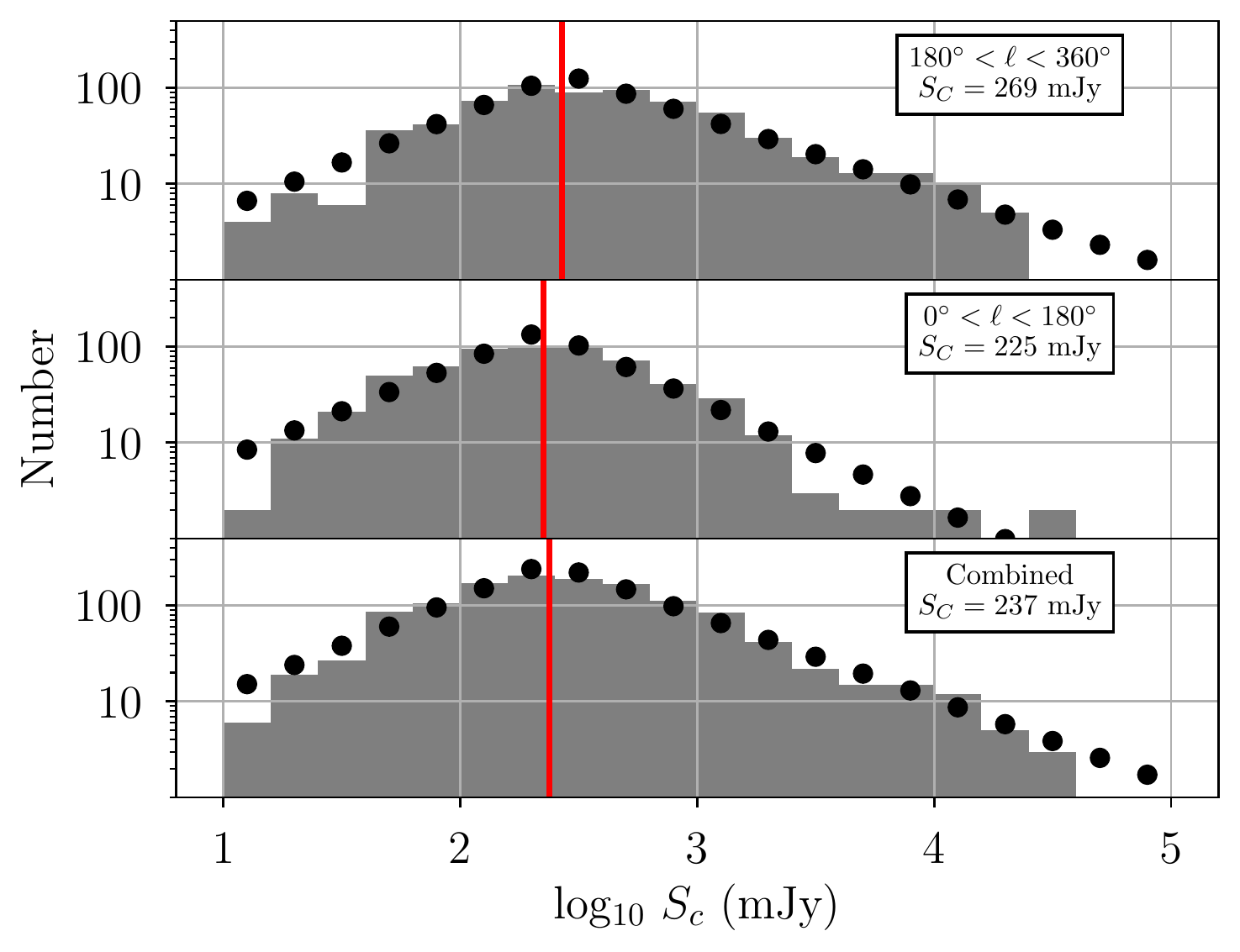}
  \caption{The distribution of GBT HRDS and SHRDS \hii\ region
    9\ghz\ continuum flux densities. The top panel shows only those
    nebulae in the third and fourth Galactic quadrants (\(180^\circ <
    \ell < 360^\circ\); \(n=728\)) and the middle panel shows only
    those nebulae in the first and second Galactic quadrants
    (\(0^\circ < \ell < 180^\circ\); \(n=606\)). The bottom panel
    shows all Galactic \hii\ regions with GBT HRDS or SHRDS flux
    density measurements \(n=1334\). The black points represent the
    expected population in each bin from the maximum likelihood power
    law distribution (Equation~\ref{eq:powerlaw}).  The vertical red
    lines are the maximum likelihood 9\ghz\ continuum flux density
    completeness limits, \(S_C\), derived for each sample:
    \(269\,\text{mJy}\) (top), \(225\,\text{mJy}\) (middle), and
    \(237\,\text{mJy}\) (bottom). This analysis suggests that the
    Galactic \hii\ region census is complete to
    \({\sim}250\,\text{mJy}\).}
  \label{fig:completeness}
\end{figure*}

\subsection{\hii\ Region Census Completeness}

We estimate the completeness of the \textit{WISE} Catalog Galactic
\hii\ region census, including the SHRDS Full Catalog data, by making
two assumptions: (1) the Galactic \hii\ region luminosity function at
\({\sim}9\ghz\) is a power law \citep[e.g.,][]{smith1989, mckee1997,
  paladini2009,mascoop2021inprep}, and (2) the nebulae are distributed
roughly uniformly across the Galactic disk
\citep[see][]{anderson2011}. Under these assumptions, the observed
flux density distribution of Galactic \hii\ regions should also follow
a power law, and we estimate the completeness of the sample as the
point where the observed flux density distribution deviates from a
power law.

We use a maximum likelihood approach to fit a power law distribution
to the continuum flux density distribution. The assumed form of the
flux density probability distribution function, \(p(S; \beta, S_c)\),
is
\begin{equation}
 p(S; \beta, S_c) = \begin{cases}
    A\left(\frac{S}{S_c}\right)^{-\beta} & S > S_c \\
    A & S_m \leq S < S_c \\
    0 & S \leq S_m
 \end{cases},\label{eq:powerlaw}
\end{equation}
where \(S\) is the flux density, \(S_c\) is the point at which the
distribution breaks from a power law, \(S_m\) is the minimum observed
flux density, and \(A\) is a normalization constant given by
\begin{equation}
  A = \frac{\beta - 1}{\beta(S_c - S_m) + S_m}.
\end{equation}

Figure~\ref{fig:completeness} shows the distribution of continuum flux
densities for Galactic \hii\ regions with GBT HRDS or SHRDS continuum
flux density measurements in three Galactic longitude zones. The top
panel is the third and fourth Galactic quadrants (\(180^\circ < \ell <
360^\circ\)), which includes the majority of the SHRDS sources. The
middle panel is the first and second quadrants (\(0^\circ < \ell <
180^\circ\)), which contains the majority of the HRDS nebulae as well
as 12 SHRDS detections. The bottom panel is the whole Galaxy. Assuming
the nebulae are optically thin, we use a thermal spectral index,
\(\alpha = -0.1\), to scale all flux densities to \(9\ghz\). The GBT
HRDS did not observe previously known nebulae. Therefore, the middle
panel in Figure~\ref{fig:completeness} has an obvious lack of bright
nebulae.

Under these assumptions, we estimate that the \textit{WISE} Catalog
Galactic \hii\ region census is complete to a \(9\ghz\) continuum flux
density limit of \({\sim}250\,\text{mJy}\). This flux density
corresponds to the brightness of an \hii\ region ionized by a single
O9 V star at \(10\,\text{kpc}\) \citep{anderson2011}. To derive this
completeness estimate, we generate 10,000 Monte Carlo realizations of
the \hii\ region data by resampling the continuum flux densities
within their uncertainties. We fit the broken power law distribution
(Equation~\ref{eq:powerlaw}) to each realization.  The median values
of the power law breaks, \(S_c\), are \(269\,\text{mJy}\) (\(180^\circ
< \ell < 360^\circ\)), \(225\,\text{mJy}\) (\(0^\circ < \ell <
180^\circ\)), and \(237\,\text{mJy}\) (combined). Assuming that the
power law breaks represent the completeness of each sample, then the
completeness in the third and fourth quadrants is \({\sim}15\%\) worse
than that of the first and second quadrants.  There are at least three
important limitations to our completeness analysis. First, the GBT
continuum flux density measurements may be inaccurate
\citep[see][]{wenger2019b}. Second, the assumption of a homogeneous
distribution of \hii\ regions across the Galactic disk is a simplified
approximation (see the following section). Third, the SHRDS continuum
flux densities may be underestimated due to missing flux.

\begin{figure*}
  \centering
  \includegraphics[width=\textwidth]{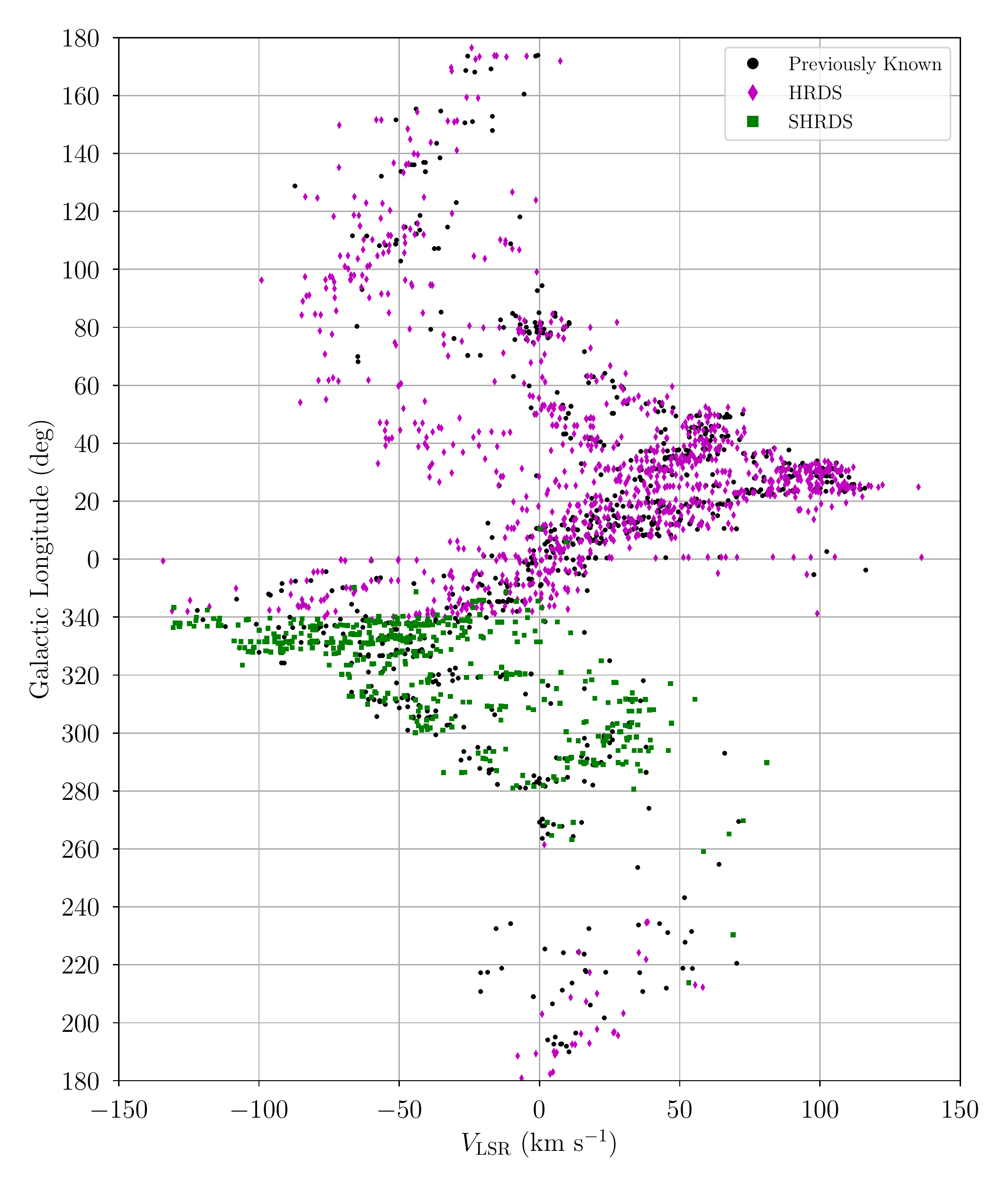}
  \caption{Galactic longitude, \(\ell\), and LSR velocity, \(V_{\rm
      LSR}\), for all known Galactic \hii\ regions in the
    \textit{WISE} Catalog with \(|V_{\rm LSR}| < 150\kms\). Including
    the \({\sim}40\) nebulae outside of this velocity range, there are
    2376 known Galactic \hii\ regions: 961 previously known nebulae
    from the \textit{WISE} Catalog (black points), 979 GBT and Arecibo
    HRDS nebulae (magenta diamonds), and 436 SHRDS nebulae (green
    squares). Some nebulae have multiple ionized gas velocity
    components; we include one data point for each component.}
  \label{fig:lv}
\end{figure*}

\begin{figure*}
  \centering
  \includegraphics[width=\textwidth]{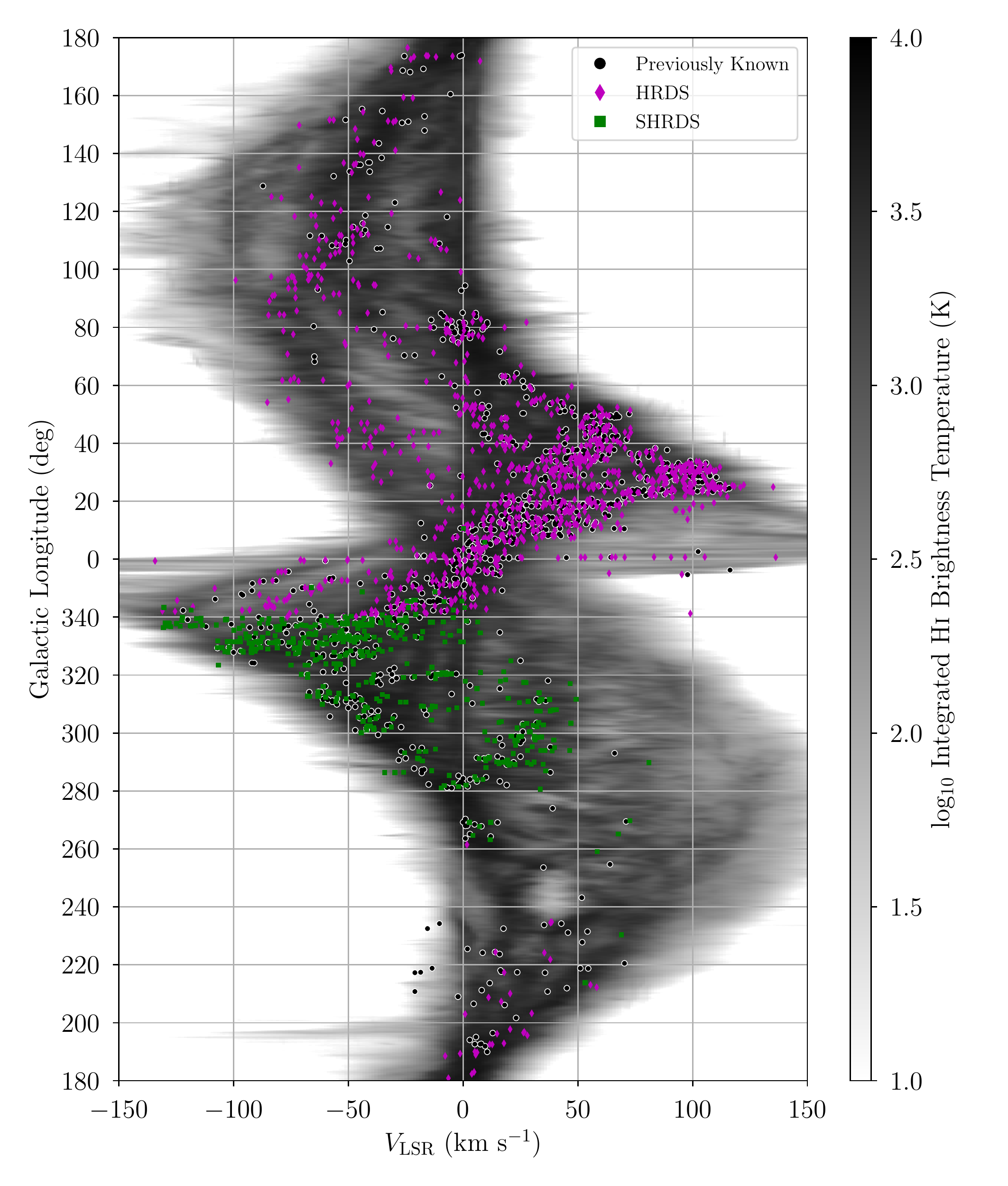}
  \caption{Same as Figure~\ref{fig:lv}, plotted atop the
    \citet{hi4pi2016} \hi\ brightness temperature integrated over
    \(|b| < 2^\circ\).}
  \label{fig:lv_hi}
\end{figure*}

\subsection{Southern vs. Northern Nebulae}

The simplest model for the distribution of Galactic \hii\ regions is a
population of nebulae distributed homogeneously across the Milky Way's
disk. This is our null-hypothesis and a useful starting point for
future explorations of more complicated models. Given a homogeneous
\hii\ region distribution and axisymmetric Galactic rotation, one
would expect a symmetric Galactic longtidue-velocity (\(\gl\)-\(v\))
distribution of \hii\ regions between the northern and southern
skies. For the first time, we use a complete census of Galactic
\hii\ regions across the entire sky to test whether such global
symmetry exists.  Figure~\ref{fig:lv} shows the \(\gl\)-\(v\)
distribution of all known Galactic \hii\ regions in the \textit{WISE}
Catalog, including those discovered in the SHRDS.
Figure~\ref{fig:lv_hi} shows the same information plotted atop a grey
scale image of the distribution of 21 cm \hi\ emission
\citep{hi4pi2016}.

\begin{deluxetable}{lccccc}
  \centering
  \tablewidth{0pt}
  \tabletypesize{\scriptsize}
  \tablecaption{Galactic \hii\ Region Census Properties\label{tab:quadrant_summary}}
  \tablehead{
    \\
    \colhead{Property} & \colhead{First Quadrant} & \colhead{Second Quadrant} & \colhead{Third Quadrant} & \colhead{Fourth Quadrant} & \colhead{Entire Galaxy} \\
    \colhead{} & \colhead{(\(0^\circ < \gl < 90^\circ\))} & \colhead{(\(90^\circ < \gl < 180^\circ\))} & \colhead{(\(180^\circ < \gl < 270^\circ\))} & \colhead{(\(270^\circ < \gl < 360^\circ\))} & \colhead{(\(0^\circ < \gl < 360^\circ\))}
  }
  \startdata
  Number of \hii\ Regions & 1176 & 155 & 90 & 955 & 2376 \\
  9\ghz\ Continuum Completeness (mJy) & \multicolumn{2}{c}{269} & \multicolumn{2}{c}{225} & 237 \\
  Median LSR Velocity (\kms) & \(39.8\) & \(-47.9\) & \(11.6\) & \(-38.0\) & \(5.0\) \\
  LSR Velocity Std. Dev. (\kms) & \(39.4\) & \(21.8\) & \(21.9\) & \(45.0\) & \(54.9\) \\
  Median RRL FWHM Line Width (\kms) & \(23.7\) & \(24.6\) & \(24.1\) & \(23.9\) & \(23.9\) \\
  RRL FWHM Line Width Std. Dev. (\kms) & \(6.7\) & \(5.5\) & \(7.6\) & \(7.0\) & \(6.8\)
  \enddata
\end{deluxetable}

The asymmetry in the \hii\ region \(\gl\)-\(v\) distribution between
the northern and southern sky is striking. This \hii\ region asymmetry
is contrasted by the \hi\ symmetry in Figure~\ref{fig:lv_hi}. \hi\ is
ubiquitous in the Milky Way \citep[e.g.,][]{kalberla2009} and thus
symmetrically fills all space in the \(\gl\)-\(v\) plane allowed by
Galactic rotation.  In contrast, Galactic \hii\ regions are neither
ubiquitous nor are they distributed homogeneously. They do not
populate the same \(\gl\)-\(v\) space as the neutral gas and their
\(\gl\)-\(v\) distribution is asymmetric between the northern (first
and second quadrants) and southern (third and fourth quadrants) sky.
Table~\ref{tab:quadrant_summary} lists the properties of the Galactic
\hii\ region census in each Galactic quadrant.  There are more
\hii\ regions in the northern sky (\(n=1331\)) compared with the
southern sky (\(n=1045\)). Furthermore, the asymmetry extends to the
inner vs. outer Galaxy. The ratio of known \hii\ regions in the fourth
quadrant to the first quadrant is 81\% whereas the ratio of third
quadrant to second quadrant nebulae is only 58\%.

A similar asymmetry is seen in the distribution of WISE Catalog
\hii\ regions and \hii\ region candidates \citep{anderson2014}. This
asymmetry cannot be fully explained by completeness differences
between these Galactic zones. The current Galactic \hii\ region census
has 315 third and fourth quadrant nebulae with 9\ghz\ continuum flux
densities less than the southern sky completeness limit, 269 mJy. If
this sample was complete to the northern sky limit, 225 mJy, then we
would expect to find an additional 54 southern sky nebulae. The
current \hii\ region census has 286 more northern sky nebulae than
southern sky nebulae.  This difference is \({\sim}5.3\) times greater
than what can be explained by the completeness disparity. The global
Galactic asymmetry in the distribution of the formation sites of
massive stars probably stems from spiral structure.

\begin{figure*}
  \centering
  \includegraphics[width=\textwidth]{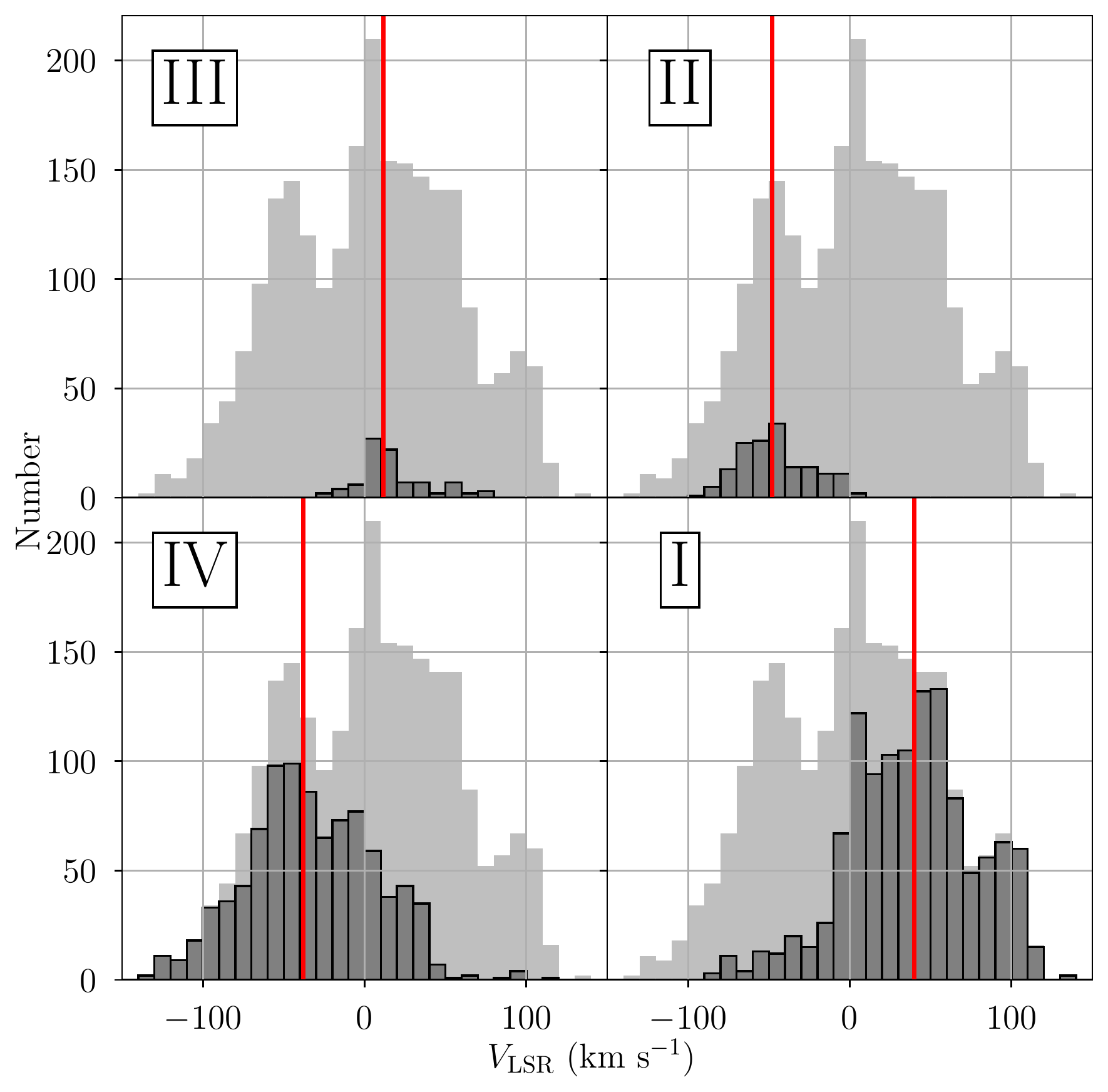}
  \caption{The distribution of \textit{WISE} Catalog \hii\ region RRL
    LSR velocities, \(V_{\rm LSR}\), in each of the Galactic
    quadrants. The median LSR velocities (number of nebulae) are
    \(39.8\kms\) (1176), \(-47.9\kms\) (155), \(11.6\kms\) (90), and
    \(-38.0\kms\) (955) in the first, second, third, and fourth
    quadrants, respectively. These median values are indicated by the
    vertical red lines in each panel. The background histrograms show
    the distribution of RRL LSR velocities for the full census of
    Galactic \hii\ regions in the \textit{WISE} Catalog. For nebulae
    with multiple RRL velocity components, only the brightest RRL
    component is included. There is an asymmetry in the \hii\ region
    LSR velocity distribution between the northern and southern sky.}
  \label{fig:quadrant_vlsr}
\end{figure*}

\begin{figure*}
  \centering
  \includegraphics[width=\textwidth]{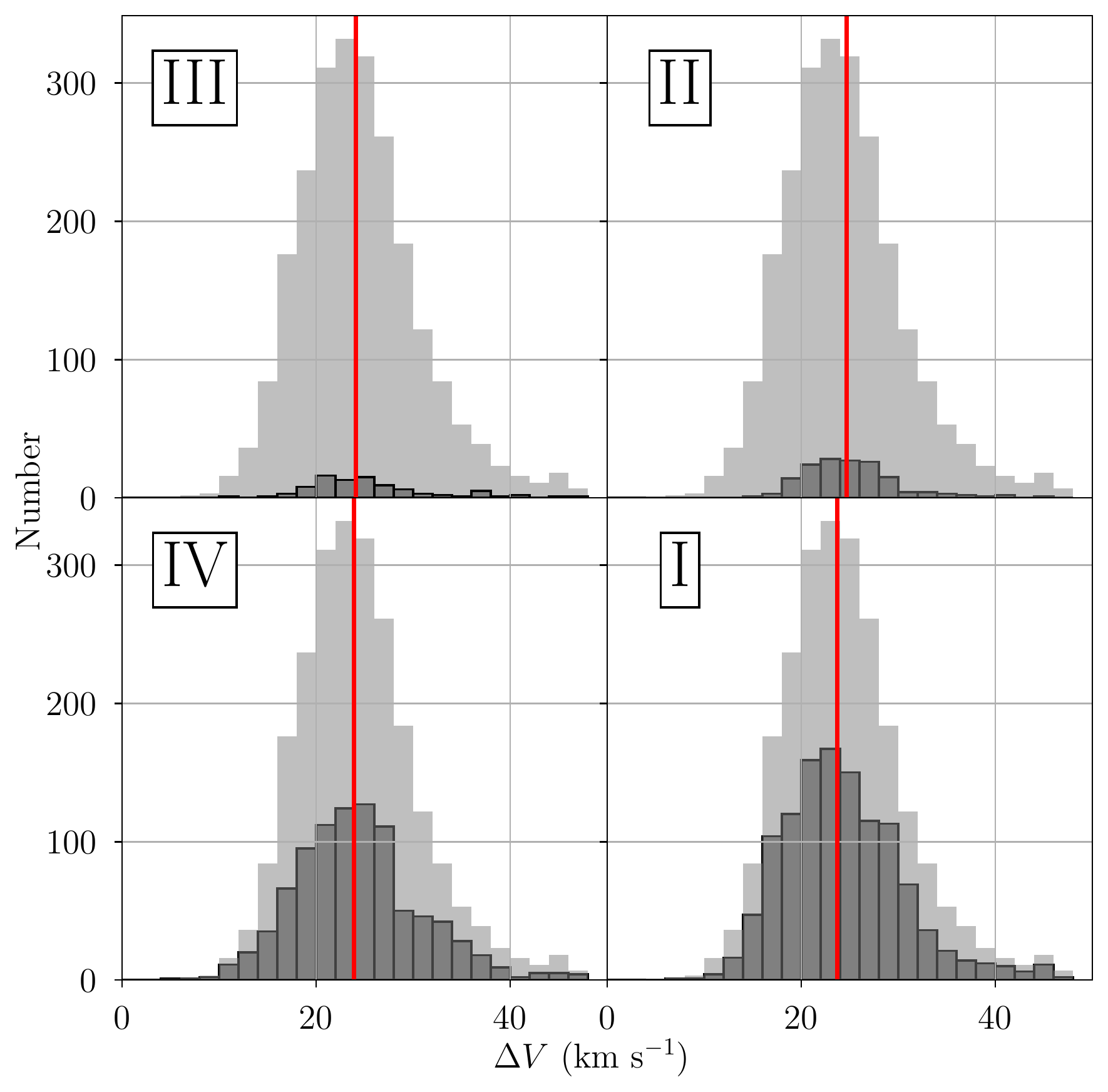}
  \caption{The distribution of \textit{WISE} Catalog \hii\ region RRL
    FWHM line widths, \(\Delta V\), in each of the Galactic
    quadrants. The median FWHM line widths are \(23.7\kms\),
    \(24.6\kms\), \(24.1\kms\), and \(23.9\kms\) in the first, second,
    third, and fourth quadrants, respectively. These median values are
    indicated by the vertical red lines in each panel.  The median
    FHWM line width of the entire population is \(23.9\kms\). The
    background histrograms show the distribution of RRL FWHM line
    widths for the full census of Galactic \hii\ regions in the
    \textit{WISE} Catalog. For nebulae with multiple RRL velocity
    components, only the brightest RRL component is included. There is
    no evidence that \hii\ region RRL line widths depend on Galactic
    location.}
  \label{fig:quadrant_fwhm}
\end{figure*}

The Figure~\ref{fig:lv} \(\gl\)-\(v\) diagram also reveals an
asymmetry in the \hii\ region velocity distribution. In the third and
fourth quadrants, for example, there are few nebulae with \(V_{\rm
  LSR} > 50\kms\). In comparison, the symmetric region of the first
and second quadrants, \(V_{\rm LSR} < -50\kms\), is thoroughly
populated. To better illustrate this asymmetry,
Figure~\ref{fig:quadrant_vlsr} shows the distribution of \hii\ region
LSR velocities in each Galactic quadrant. The median and standard
deviation of each distribution is listed in
Table~\ref{tab:quadrant_summary}. Over the entire Galaxy, the median
\hii\ region LSR velocity is \(5.0\kms\) with a standard deviation of
\(54.9\kms\).

The first and fourth quadrant distributions are symmetric; the
magnitudes of their median LSR velocities are within \(2\kms\) and the
widths of the distributions are comparable to within 15\%.  The median
velocities, however, are not symmetric between the second and third
quadrants. The second and third quadrants also have similar LSR
velocity distribution widths, although they are \({\sim}50\%\)
narrower than the first and fourth quadrant distributions. Such
asymmetry is not expected for a homogeneous and axisymmetric
distribution of \hii\ regions. This velocity asymmetry is probably an
artifact of spiral structure in the \hii\ region distribution.

The RRL FWHM linewidth distribution of the southern sky nebulae is
nearly indistinguishable from the northern sky
population. Figure~\ref{fig:quadrant_fwhm} shows the distribution of
RRL FWHM linewidths in each of the Galactic quadrants. The median and
standard deviation of each distribution is listed in
Table~\ref{tab:quadrant_summary}. The median RRL FWHM linewidth of all
known Galactic \hii\ regions in the \textit{WISE} Catalog, including
the new SHRDS nebulae, is \(23.9\kms\) with a standard deviation of
\(6.8\kms\). Such uniformity of these distributions implies that the
physical conditions of \hii\ regions (e.g., internal turbulence) must
be similar across the Galaxy.

\section{Completing the Census of Galactic \hii\ Regions}

With this data release, we conclude our \hii\ Region Discovery
Surveys, which began with the GBT HRDS nearly a decade ago. The SHRDS
Full Catalog data are included both in the latest release of the
\textit{WISE} Catalog of Galactic \hii\ Regions \citep{anderson2014}
and in a machine-readable database of ionized gas detections toward
Galactic
\hii\ regions\footnote{\url{https://doi.org/10.7910/DVN/NQVFLE}}. The
\textit{WISE} Catalog now contains 2376 known \hii\ regions, 1690
candidates, 3718 radio-quiet candidates, and 632 group candidates.
The \hii\ region census is complete to \({\sim}250\) mJy at 9\ghz. The
remaining \hii\ region candidates are likely distant or ionized by
lower mass stars \citep{armentrout2021}.

With \({\sim}6000\) remaining \hii\ region candidates in the
\textit{WISE} catalog, the census of Galactic \hii\ regions is still
incomplete. New single dish RRL surveys in search of new Galactic
\hii\ regions must be mindful that poor angular resolution complicates
the interpretation of the source of the emission. For example,
\citet{chen2020} claim the detection of RRL emission in \({\sim}75\)
directions not associated with a \textit{WISE} Catalog object. An
inspection of these positions reveals that many are overlapping or
immediately adjacent to one or more \textit{WISE} Catalog sources.
The source of the detected RRL emission could therefore be the
\hii\ region itself, ionized gas in the immediate vicinity of the
\hii\ region due to leaked ionizing photons
\citep[e.g.,][]{luisi2016}, or diffuse ionized gas (DIG) somewhere
along the line of sight \citep[e.g.,][]{liu2019}. As we push to better
sensitivity in single dish surveys, it will become more challenging to
disentangle RRL emission toward discrete sources from the DIG.

To complete the census of Galactic \hii\ regions and find RRL emission
toward the remaining \textit{WISE} Catalog candidates, the next
generation of \hii\ region discovery surveys will require better
sensitivity. A nebula ionized by a single B0 V star 25 kpc distant,
for example, has a 9\ghz\ continuum flux density of
\({\sim}10\,\text{mJy}\)
\citep{anderson2011,armentrout2021,mascoop2021inprep}. Assuming
a typical RRL-to-continuum ratio of \(0.1\) at \(9\ghz\), the RRL flux
density is \({\sim}1\,\text{mJy}\). The VLA would require an on-source
integration time of \({\sim}30\,\text{min}\) (\(560\,\text{min}\) for
the ATCA) to achieve a \(5\sigma\) RRL detection in \(5\kms\) channels
after averaging 8 RRL transitions. These integration times are
impracticable for a survey of several thousand \hii\ region
candidates. Future facilities, such as the Next Generation Very Large
Array (ngVLA) and Square Kilometer Array (SKA) will be capable of
finding RRLs toward most, if not all, of the remaining Galactic
\hii\ region candidates. The angular resolution of these instruments
will also allow for the extinction-free detection of discrete
\hii\ regions in nearby galaxies \citep{balser2018}.

\section{Summary}

The SHRDS comprises the final contribution to our \hii\ Region
Discovery Surveys. With the ATCA, we detect 4--10\ghz\ radio continuum
emission toward 212 previously known \hii\ regions and 518
\hii\ region candidates. We detect RRL emission toward 208 previously
known nebulae and 438 \hii\ region candidates by averaging
\({\sim}18\) RRL transitions. The detection of RRL emission from these
nebulae thus increases the number of known \textit{WISE} Catalog
Galactic \hii\ regions in the surveyed zone by 130\% to 778
nebulae. Including the previous northern sky surveys with the GBT and
Arecibo Telescope, as well as the SHRDS, the HRDS has now discovered
\({\sim}1400\) new nebulae. These discoveries made during the past
decade have more than doubled the number of known Galactic
\hii\ regions in the \textit{WISE} Catalog.

All SHRDS data products, including continuum images and data cubes for
each RRL transition, are publicly available\footnote{See
  \url{https://www.cadc-ccda.hia-iha.nrc-cnrc.gc.ca/en/community/shrds/}. The
  SHRDS data products are archived at the Canadian Advanced Network
  for Astronomical Research (\url{https://doi.org/10.11570/21.0002})
  and the CSIRO Data Access Portal in Australia (the first of fifteen
  data groups is at \url{https://doi.org/10.25919/7nf1-n140)}.}. These
data are included both in the latest release of the \textit{WISE}
Catalog of Galactic \hii\ Regions \citep{anderson2014} and in a
machine-readable
database\footnote{\url{https://doi.org/10.7910/DVN/NQVFLE}}. A single
nebula may appear multiple times in our radio continuum and RRL
catalogs because we analyze the data in several different ways. For
example, a source may be detected in both the non-smoothed and
smoothed images. For each detected nebula we measure both the peak and
total continuum flux density. We determine the RRL properties in
spectra extracted from the brightest pixel as well as in spectra
averaged over all pixels containing emisssion. Furthermore, a nebula
may be detected in multiple individual fields or mosaics. Users of
these catalogs should use the data that best fit their science goals.

The census of Galactic \hii\ regions in the \textit{WISE} Catalog is
now complete to \({\sim}250\,\text{mJy}\) at \(9\ghz\). This flux
density is equivalent to a nebula ionized by a single O9 V star at a
distance of \({\sim}10\,\text{kpc}\). The distribution of \hii\ region
RRL line widths is similar in each Galactic quadrant, with a median
FWHM line width of \(23.9\kms\). The asymmetry in the number of
nebulae and the distributions of RRL LSR velocities probably stems
from Galactic spiral structure.

With a flux-limited sample of nebulae across the Milky Way, we can now
begin to craft a comprehensive view of Galactic chemical and
morphological structure. A face-on map of the \hii\ region locations
requires accurate distances, which can be derived from maser parallax
measurements \citep[e.g.,][]{reid2019} or kinematically
\citep[e.g.,][]{wenger2018}. In an upcoming paper in this series, we
will use \hi\ absorption observations to resolve the kinematic
distance ambiguity and thus derive the distances to hundreds of SHRDS
nebulae.

The next generation of Galactic \hii\ region discovery surveys will
require the sensitivity of future facilities, such as the ngVLA and
the SKA. These telescopes will have the sensitivity to detect
continuum and RRL emission toward most Galactic \hii\ regions as well
as the angular resolution to resolve the emission from discrete
\hii\ regions in nearby galaxies. Such extragalactic studies will not
only provide an extinction-free tracer of star formation in these
galaxies, but also a direct comparison to what we are learning about
Milky Way high-mass star formation, Galactic chemical structure, and
Galactic morphological structure.

\acknowledgments

We thank the anonymous referee for their feedback, which improved the
clarity of this paper. L.D.A. is supported in part by NSF grant
AST-1516021. T.M.B. is supported in part by NSF grant
AST-1714688. J.R.D. is the recipient of an Australian Research Council
(ARC) DECRA Fellowship (project number DE170101086). This research has
made use of NASA's Astrophysics Data System Bibliographic Services. We
are grateful for the storage space and computing resources provided by
the NRAO for this experiment as well as the archive storage space
provided by the Canadian Advanced Network for Astronomical Research
(CANFAR) and CSIRO.

The National Radio Astronomy Observatory and Green Bank Observatory
are facilities of the National Science Foundation operated under
cooperative agreement by Associated Universities, Inc. The Australia
Telescope Compact Array is part of the Australia Telescope National
Facility, which is funded by the Australian Government for operation
as a National Facility managed by CSIRO.

\facility{ATCA}

\software{Astropy \citep{astropy2013},
  CASA \citep{mcmullin2007},
  Matplotlib \citep{matplotlib2007},
  NumPy \& SciPy \citep{numpyscipy2011},
  WISP \citep{wisp}}

\bibliography{shrds_fullcatalog}

\appendix

\section{Sinc Interpolation\label{app:sinc}}

Sinc interpolation provides a perfect reconstruction of discrete
signals sampled at the Nyquist rate or higher
\citep{oppenheim1975}. It is the proper way to interpolate
time-limited (or band-limited), discrete frequency (or time) signals
to arbitrary frequencies (or times) and/or sampling rates. For an
original spectrum, \(s(v)\), sampled at velocities, \(v_0, v_1, ...,
v_N\), the smoothed and re-sampled spectrum, \(s^*(v^*)\), sampled at
velocities, \(v^*_0, v^*_1, ..., v^*_M\), with channel spacing
\(\Delta v^*\) is the linear convolution of \(s(v)\) with a sinc
function,
\begin{equation}
  s^*(v^*_j) = \sum_i^Ns(v_i)A_j\,\text{sinc}\left(\frac{v^*_j - v_i}{\Delta v^*}\right),
\end{equation}
where \(A_j\) is a normalization constant such that
\begin{equation}
  \sum_i^NA_j\,\text{sinc}\left(\frac{v^* - v_i}{\Delta v^*}\right) = 1,
\end{equation}
and \(\text{sinc}(x) = \sin(\pi x)/(\pi x)\).

Smoothing and interpolating large data cubes is implemented
efficiently by treating the convolution as a tensor dot product. For a
multi-dimensional data cube \(\vect{S}\) with the \(N\)-length
velocity axis along the first dimension (e.g., a four-dimensional data
cube with the typical radio astronomy axes: velocity, Galactic
latitude, Galactic longitude, and Stokes), the smoothed data cube with
the \(M\)-length re-gridded velocity axis along the first dimension is
the tensor dot product,
\begin{equation}
  \vect{S}^* = \vect{W}\cdot\vect{S},
\end{equation}
where \(\vect{W}\) is the \(M\times N\) array of sinc weights with
elements
\begin{equation}
  w_{ij} = A_j\,\text{sinc}\left(\frac{v^*_j - v_i}{\Delta v^*}\right).
\end{equation}

\end{document}